\documentclass[format=acmsmall, nonacm, screen=True]{acmart}
\pdfoutput=1 
\setcopyright{none}
\AtBeginDocument{%
  \providecommand\BibTeX{{%
    \normalfont B\kern-0.5em{\scshape i\kern-0.25em b}\kern-0.8em\TeX}}}

\usepackage{booktabs} 
\usepackage[ruled]{algorithm2e} 
\usepackage{tabularx}

\SetAlFnt{\small}
\SetAlCapFnt{\small}
\SetAlCapNameFnt{\small}
\SetAlCapHSkip{0pt}
\IncMargin{-\parindent}

\usepackage{makecell}
\usepackage{lscape} 
\usepackage{float}
\usepackage{threeparttable}  
\usepackage{booktabs}
\usepackage{multirow}
\setcopyright{acmcopyright}
\copyrightyear{2018}
\acmYear{2018}
\acmDOI{10.1145/1122445.1122456}

\graphicspath{{figures/}}

\usepackage{lipsum}
\newcommand{\beginsupplement}{%
        \setcounter{table}{0}
        \renewcommand{\thetable}{S\arabic{table}}%
        \setcounter{figure}{0}
        \renewcommand{\thefigure}{S\arabic{figure}}%
     }
\newcommand{\rev}[1]{{\color{black}#1}}

\begin{document}
\title[Concerns and Value Judgments of Stakeholders in the Non-Fungible Tokens (NFTs)
Market]{"Centralized or Decentralized?": Concerns and Value Judgments of Stakeholders in the Non-Fungible Tokens (NFTs) Market}

\author{Yunpeng Xiao}
\affiliation{\institution{Illinois Institute of Technology}\city{Chicago}\state{Illinois}\country{United States}}

\author{Bufan Deng}
\affiliation{\institution{Technical University of Denmark}\city{Copenhagen}\country{Denmark}}

\author{Siqi Chen}
\affiliation{\institution{University of Texas, Austin}\city{Austin}\state{Texas}\country{United States}}

\author{Kyrie Zhixuan Zhou}
\affiliation{\institution{University of Illinois at Urbana-Champaign}\city{Champaign}\state{Illinois}\country{United States}}

\author{RAY LC}
\affiliation{\institution{School of Creative Media, City University of Hong Kong}\city{Hong Kong}\country{China}}

\author{Luyao Zhang}
\affiliation{\institution{Duke Kunshan University}\city{Suzhou}\state{Jiangsu}\country{China}}

\authornote{Corresponding authors: \newline
Luyao Zhang (email: lz183@duke.edu) and Xin Tong (email: xt43@duke.edu), address: Duke Kunshan University, No.8 Duke Ave. Kunshan, Jiangsu 215316, China.}

\author{Xin Tong}
\affiliation{\institution{Duke Kunshan University}\city{Suzhou}\state{Jiangsu}\country{China}}
\authornotemark[1]

\begin{abstract}
Non-fungible tokens (NFTs) are decentralized digital tokens to represent the unique ownership of items. Recently, NFTs have been gaining popularity and at the same time bringing up issues, such as scams, racism, and sexism. Decentralization, a key attribute of NFT, contributes to some of the issues that are easier to regulate under centralized schemes, which are intentionally left out of the NFT marketplace. In this work, we delved into this centralization-decentralization dilemma in the NFT space through mixed quantitative and qualitative methods. Centralization-decentralization dilemma is the dilemma caused by the conflict between the slogan of decentralization and the interests of stakeholders.  We first analyzed over 30,000 NFT-related tweets to obtain a high-level understanding of stakeholders' concerns in the NFT space. We then interviewed 15 NFT stakeholders (both creators and collectors) to obtain their in-depth insights into these concerns and potential solutions. Our findings identify concerning issues among users: financial scams, counterfeit NFTs, hacking, and unethical NFTs. We further reflected on the centralization-decentralization dilemma drawing upon the perspectives of the stakeholders in the interviews. Finally, we gave some inferences to solve the centralization-decentralization dilemma in the NFT market and thought about the future of NFT and decentralization.

\end{abstract}



\maketitle
\section{Introduction}
Non-Fungible Tokens (NFTs) are blockchain-based digital assets that represent physical or digital creative work, and intellectual property of music, digital art, games, gif pictures, video clips, etc.~\citep{rehman2021nfts}. In the first half year of 2021, the NFT market generated around \$24.7 billion worth of organic trading volume across blockchain platforms and marketplaces~\citep{NFTsale2021}. For instance, Beeple's NFT artwork was sold for a record-high 69 million USD in March 2021~\citep{tarkha2021}. 
\
\newline
\indent
However, some issues cannot be ignored in the NFT market despite its rapid development. Previous studies point to some issues in the NFT market, such as financial scams ~\cite{sharma2022s,das2022understanding}, financial bubbles~\citep{chalmers2022beyond}, hacking~\citep{das2022understanding,bamakan2022patents,kshetri2022scams}, counterfeit NFTs~\citep{bamakan2022patents,das2022understanding}, and inequality and discrimination ~\citep{zhang2022visualizing,nguyen2022racial,zhang2023mechanics}. In addition to these issues, there are even more concerns perceived by people in the wider blockchain field, including philosophical significance, environmental protection, accessibility, public governance, and privacy protection~\citep{tang2019ethics}. 

Compared to traditional E-commerce platforms such as Amazon and Taobao, NFTs are traded with blockchain. Decentralization is the core concept of blockchain. It is known to prevent authority abuse and corruption arising from centralized regulation~\citep{zheng2018blockchain}, such as governments and banks. However, some studies pointed out decentralization also introduces issues that may endanger the rights of community members~\citep{zhang2019security}. Some scholars also thought that security vulnerabilities, such as financial scams, and some ethical concerns, such as racism/sexism~\citep{zhang2022visualizing}, can be more effectively solved under centralized regulation ~\citep{liu2017centralization}. Some studies have further pointed out that the decentralization of technology does not necessarily lead to substantial decentralization~\citep{golumbia2016politics, ao2023decentralized, zhang2022blockchain,zhang2022sok,zhang2023design}. Decentralization is not only a technical concept but has already become the slogan of the blockchain community~\citep{golumbia2016politics}. Centralized regulation is obviously against this slogan.
\
\newline
\indent
At present, many scholars have considered the decentralization of the blockchain from the perspective of economics and sociology~\citep{de2020blockchain, atzori2015blockchain, schneider2017once, golumbia2016politics}, but the research on the NFT market is still relatively limited. Moreover, these studies often rely on logical inferences and rarely have the perspective of users. People's perceptions and understanding lead them to make value judgments in the NFT marketplace, thus it is crucial to understand their views. Proponents of decentralization claim that in the absence of a centralized institution, every member of a decentralized community will participate in community governance ~\citep{atzori2015blockchain}. Thus, if the needs and experience of members in the NFT market are not well understood, the designed technologies and formulated rules may be biased, which is not conducive to the further development of the NFT market. 

In the current investigation, we focus on NFT creators and collectors and propose the following research questions:
\par• \textbf{RQ1:} What are the perceptions and concerns of stakeholders in the NFT markets?
\par• \textbf{RQ2:} Is centralization the panacea for solving these issues in the NFT markets? 

We conducted Twitter analysis and interviews. In the Twitter analysis, to investigate the issues of greatest concern to stakeholders, we constructed a network graph of social networks and found Key Opinion Leaders (KOLs); Then, we randomly sampled 250 tweets from KOLs and 200 tweets from common users for a qualitative analysis. We found users were most concerned about financial scams. Other issues in the NFT market included counterfeit NFTs, hacking, and unethical NFTs. However, the discussion on Twitter is not deep enough, and people rarely talk about decentralization and solutions to these issues, such as whether centralized regulation should be adopted. Therefore, we recruited 15 participants for interviews by posting information on Twitter and Weibo and snowball sampling. During the interview, we asked about participants' perceptions of decentralization. We found that participants differed widely on whether to support centralized regulation on different issues. Many people were hesitant to use centralized regulation. We called this phenomenon the "Centralization-Decentralization Dilemma". Its essence is the dilemma caused by the conflict between the slogan of decentralization and the interests of stakeholders. In addition, people's perspectives on an issue are often related to their interests. Finally, we combine Twitter analysis results and interview results and propose our recommendations for solving the centralization-decentralization dilemma, such as incorporating centralized regulation to destructive issues and encouraging decentralized autonomy for ambiguous issues. It can be seen that in our methods, qualitative Twitter analysis is based on quantitative Twitter analysis, while interviews are based on Twitter analysis, and the relationship between the three is progressive. Therefore, in the results section (section 4), we list the results obtained by the three methods in order and finally answer our RQs in section 4.4.

\section{Related Work}
\subsection{Introduction to NFTs}
Non-Fungible Tokens (NFTs) are digital assets that represent the ownership of objects. They cannot be exchanged like-for-like or subdivided. In the orthodox sense, the objects contained in NFTs are digital assets~\citep{wang2021non}, creative work, or intellectual property~\citep{rehman2021nfts}. However, they represent not only the ownership of virtual assets but also real assets~\citep{das2022understanding}. The trading of NFTs is similar to traditional commodity (or goods) trading. Compared to traditional E-commerce platforms such as Amazon and Taobao, NFTs are traded with blockchain instead. \rev{A general NFT consists of two parts: the token representing ownership and the corresponding asset. Tokens are stored in the blockchain, and assets have different storage methods. For example, digital assets can be stored in the interplanetary file system (IPFS)~\citep{psaras2020interplanetary}. And for real assets, It is impossible for them to be stored in computers.}

\subsubsection{Technical Components of NFTs}
\ 
\newline
Hereby we introduce some technical details about NFTs, 
particularly relating to the decentralization issues.

\textbf{Blockchain and Cryptocurrency}
One of the most important issues of centralized regulation is that authoritative organizations or individuals can tamper with records at will. Blockchain was first proposed in 1991~\citep{haber1990time}. Most blockchains use the proof of work (PoW) mechanism, where all full nodes (so-called miners) compete for the authority to create blocks by computing block hashes. When a miner obtains the authority to create blocks, it can package transactions into a block and broadcast it to the rest of the miners. After other miners verify the legitimacy of this block (by calculating the block hash), they immediately stop calculating the current block hash and start calculating the next block hash. Transactions are saved in the data structure of a Merkle tree~\citep{merkle2019protocols}. Once a transaction is tampered with, the hash value of the Merkle tree root node will change, and the hash value of the entire block will also change. So, it is hard to tamper with the transaction records. Also, the transaction records are public, and anyone can see transaction records if they want. Correspondingly, this verification mechanism is very inefficient, and when the transaction volume is too large, the transaction waiting time will be very long.~\citep{liu2022empirical,zhang2023understand,heimbach2023defi}

NFTs are mainly traded through the Ethereum blockchain~\citep{wood2014ethereum}. Ethereum has been using the PoW mechanism for the past seven years, but starting from September 15, 2022, Ethereum began to use the Proof of Stake (PoS) mechanism. This is because a large number of hash operations of the PoW mechanism has caused environmental problems that cannot be ignored~\citep{kapengut2022event, zhou}. However, PoS also brings new security issues~\citep{schwarz2022three}. There are also other blockchains to create or trade NFTs, such as Solana, which uses a Proof of History mechanism~\citep{yakovenko2018solana}. 

\textbf{Address, Smart Contract, Transaction, and Verification}
The blockchain address is the unique identifier that users use to pay and receive assets. It contains a pair of randomly generated public and private keys~\citep{wang2021non}. In transactions, the users must prove their identities and use private keys to create corrected digital signatures, and full nodes need to use the public keys to verify the digital signatures. A transaction is legitimate if all digital signatures are verified correctly. In NFT transactions, in addition to verifying digital signatures, smart contracts also need to be verified. A smart contract is a computer program consisting of a set of rules~\citep{wang2018overview}. Effective proof mechanisms, like PoW, ensure that the contract execution results will not be tampered with. Users need to pay gas fees according to the complexity of the smart contracts to block creators. Smart contracts enable users to use more diverse forms, such as auctions~\citep{hong2019design}, to conduct transactions. 

\textbf{Forks in the blockchain}
A fork in a blockchain means that the blockchain diverges into two paths forward. Forks are usually caused by changes in the blockchain protocol. Forks can be categorized into hard forks and soft forks. Hard forks are permanent forks in the system~\citep{buterin2017hard}, which make previously invalid blocks and transactions valid. After hard forks, miners have to upgrade their clients, otherwise, they cannot stay in the upgraded hard fork.

\subsubsection{NFT Marketplaces}
\ 
\newline
\indent
\textbf{Transaction process of NFTs}
An NFT marketplace is mainly composed of two parts: a web front end that interacts with users and a collection of smart contracts that interact with the blockchain. Users interact with the web front, which sends transactions to the smart contracts on their behalf~\citep{das2022understanding}. Each transaction incurs gas fees and marketplace fees. Marketplace fees vary greatly among different marketplaces. The gas fees of different blockchains are also different. Ethereum's gas fees average 70 dollars, while in Solana it is only 0.01 dollars~\citep{NFTcost2022}. 

\begin{figure}[H]
     \centering    
     \includegraphics[width=1.0\textwidth]{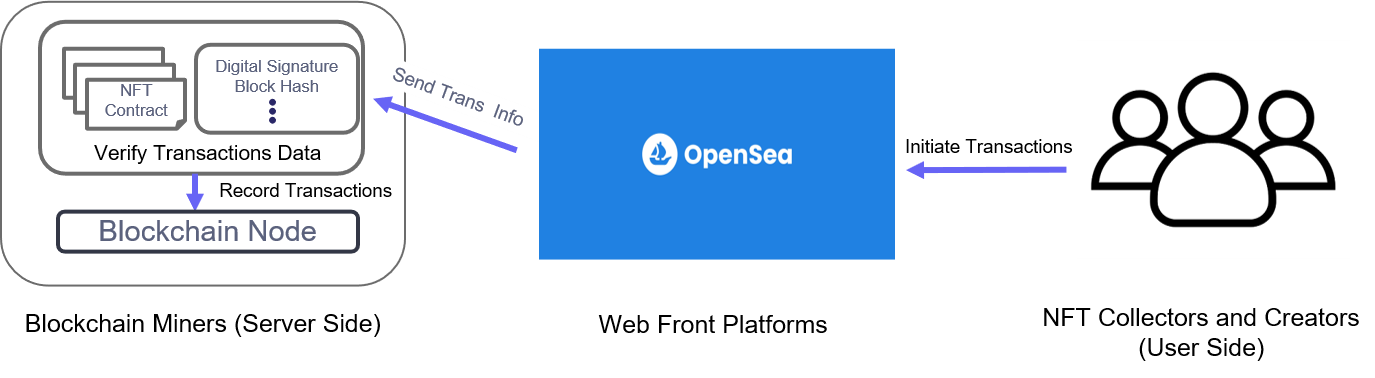}
    \caption{NFT transaction process. Creating NFT (also known as NFT minting) is a special transaction of NFT.}
     \label{fig:Transaction_Process}
 \end{figure}

A new NFT first needs to be minted on the blockchain. Minting is a special transaction, which also encodes into blockchains and incurs gas fees and marketplace fees. The people who mint NFT(s) are called NFT creators. It should be emphasized that the web front of marketplaces is highly centralized~\citep{bamakan2022patents}. The simple transaction process is shown in Figure.~\ref{fig:Transaction_Process}.
\ 
\newline
\indent
\textbf{History and stakeholders of NFT marketplaces}
The concept of NFT was first proposed in 2014~\citep{FirstNFT2014}. However, this concept is only widely used after the ERC-721~\citep{entriken2018eip} annotation was proposed in 2017. In 2021, NFT sales started to grow explosively~\citep{NFTincrease2022}. NFT market transactions are flatlining from May 2022~\citep{NFTflatlining2022}. Stakeholders in the NFT market are generally considered to be creators and collectors. However recent work has divided stakeholders into six categories in greater detail~\citep{baytacs2022stakeholders}.
\ 
\newline
\indent
\textbf{NFT community}
Community plays a crucial role in the NFT ecosystem. Blockchain enthusiasts believe that decentralized autonomous organizations (DAOs) have the potential to enhance democracy and create new models for community-driven support of arts ~\citep{whitaker_art_2019}. 
By leveraging the communities, NFT projects are able to maintain community engagement and facilitate easy onboarding~\citep{kaczynski2021nfts}. 
NFT communities widely exist on Twitter, NFT subreddits, Discord channels of NFT projects, etc. \cite{twitter:1,twitter:2}. Among them, the community of Bored Ape Yacht Club is the most thriving one \cite{bored:ape, casale2022impact}.
Moderators
are widely introduced in community-based social platforms such as Discord.
However, it is hard to differentiate between hyping and a real commitment to community values. Hot slangs, e.g., HODL and WGNMI, encourage investors to buy-and-hold indefinitely, bidding higher values. Catlow et al.~\citep{catlow2017artists} were concerned that the frequent use of revolutionary rhetoric and feel-good language in social organizations could lead to a bubble of `The Madness of Crowds' as in the Tulip Mania of the Dutch Golden Age\footnote{A period in the Dutch Golden Age during which the prices of tulip bulbs skyrocketed, leading to an economic bubble that eventually burst.}.

\subsection{Decentralization Issues in Blockchain and NFT}
Since the blockchain technology was born in 2008, the debate on centralization and decentralization has never stopped. Some controversies are related to blockchain technology applications, while others are closely related to social influence in general.

\subsubsection{Issues with Decentralized Applications}
\ 
\newline
In this section, we will briefly introduce the issues arising from the decentralized blockchain technology and its applications, NFT included. 

\ 
\newline
\indent
\textbf{Hacking and privacy}
\rev{As mentioned in Section 2.1, NFT transactions must go through the front-end platform and the back-end blockchain. Therefore, hacking can occur on both the front-end platform and the back-end blockchain. The hacking that happens on the front end is similar to Internet hacking. The most common situation is to steal users' private keys or assets through phishing websites~\citep{aleroud2017phishing}. Decentralization increases the risk of losing assets or private keys:} Although decentralization makes it difficult to tamper with transactions, it also means that once a user's assets or private keys are lost or stolen by adversaries, they are unlikely to be recovered through disputes in a centralized manner (e.g., through a bank)~\citep{park2017blockchain}. Another common hack is by exploiting smart contract vulnerabilities or creating malicious contracts~\citep{atzei2017survey, vulnerability}, they happen in the blockchain on the back-end. One well-known example is the DAO event. In 2015, the first decentralized autonomous organization, a.k.a. the DAO, was born on Ethereum. The DAO is essentially a smart contract, and its purpose, as its name suggests, is to challenge traditional centralized organizations~\cite{quan2023decoding}. However, in June 2016, Hackers exploited smart contract bugs and stole around 11 million ETH from the DAO~\citep{wang2019decentralized}. At this time, sharp divisions emerged within the Ethereum community. Some members of the Ethereum community supported the method of rolling back transactions to retrieve lost property, while others believe that this is a centralized practice that violates the "code is law" principle. "Code is law" was proposed long before the advent of information technology, but now it has become the slogan of radical supporters of blockchain and decentralization~\citep{de2018blockchain,zhang2023machine}. In the end, Ethereum underwent a hard fork to form two separate cryptocurrencies, namely, Ethereum (ETH) and Ethereum Classic (ETC)~\citep{atzei2017survey}. 

Regarding privacy, although the user's virtual address is used in the transaction (pseudo-anonymity), once the virtual address is associated with the real identity, all the user's transactions will be exposed because the transactions are completely public on the blockchain~\citep{zhang2019security}. 
\ 
\newline
\indent
\textbf{Counterfeit NFTs and controversial NFTs}
NFT minting is essentially a process of data writing. Blockchain itself cannot judge whether the data written into the chain is moral or not; it can only prevent the written data from being tampered with. Therefore, counterfeit NFTs are rampant. Common counterfeiting practices include minting other people's non-NFT creations into NFTs, or making minor modifications to existing NFTs and selling them at lower prices. 

There are also controversial NFTs minted. The meta slave project in early 2022 has caused massive controversy, and it has been accused of blatantly expressing racism against black people. This project was later removed from Opensea, a centralized NFT marketplace~\citep{Metaslave2022}. 
Such platforms have the authority to remove or push specific NFTs~\citep{NFTplatformcentralised}. This brings up a dilemma: if we want to keep users from buying counterfeit or controversial NFTs, we must rely on the authority of centralized platforms, which is what blockchain and NFT proponents are against. Existing research simply described or quantified such phenomena without considering their broader social implications~\citep{das2022understanding,rehman2021nfts}. 

\subsubsection{Debates of Blockchain and Decentralization in Social Influence}
\ 
\newline
For blockchain proponents, decentralization is not just a technical concept, but a social and philosophical slogan, sometimes even a political philosophy. This idea is closely related to the time when the blockchain was born. In 2008, the financial crisis swept the world. A large number of bank failures including bad debts~\citep{nofer2017blockchain} have triggered a crisis of trust toward centralized financial institutions (e.g. banks, stock exchanges) and centralized regulators (e.g., governments)~\citep{earle2009trust}. The emergence of the blockchain seemed to provide a solution for all these issues: since every transaction on the blockchain can be clearly and accurately recorded, it seemed that there would be no opaque and dark transactions leading to bad debts. The invention of Ethereum and the emergence of smart contracts also gave some people the belief that all behaviors on the blockchain can be regulated by code and that the role of centralized governments and laws is no longer necessary. 

The concept of decentralization has caused a lot of controversy. Some left-wing scholars believe that ``decentralization'' is just a new term for right-wing extremism. Despite the rhetoric of revolution and ``democratization,'' the essence of such rhetoric is to further strengthen the authority of capital in order to evade legal and democratic supervision. Its purpose is not ``decentralization,'' but to eliminate the centralized government and establish ``centralized capital''~\citep{golumbia2016politics}. This kind of thinking is essentially a combination of blockchain and anarcho-capitalism (a type of neoliberalism), and some scholars call it crypto-libertarians~\citep{husain2020political}. Many scholars also believe that the decentralization of technology does not mean the actual decentralization~\citep{schneider2017once, atzori2015blockchain, dodd2018social}. In the blockchain and NFT market, some notorious financial scams, such as Ponzi schemes, rug pulls~\citep{rugpull}, and wash trades~\citep{wash}, are typical manifestations of capital evading the law and democratic supervision. In addition, the high gas fee is also one of the consequences of "centralized capital." The high gas fees of some blockchains, such as Ethereum, are still a hot topic in the NFT market~\citep{gasfees}. Although it can limit the counterfeiting of NFTs to a certain extent, it may also cause creators who are not financially rich to be discouraged from NFTs~\citep{chalmers2022beyond}, making NFT a less inclusive marketplace. There are also some other theories like techno-utopianism, whose slogan is ``Code is law,'' and whose purpose is to build technical infrastructure to replace the functions of centralized organizations~\citep{tapscott2016blockchain, kshetri2017will}. This theory has similarities to crypto-libertarians but places more emphasis on the role of technology. Some scholars analyzed the DAO event and believed in that event: most people supported the use of rollback to retrieve property, which proved the failure of the techno-utopianism~\citep{hutten2019soft}. One of the reasons for the perceived failure of "code is law" is that smart contracts are implemented by humans (programmers), and it is impossible to consider all situations~\citep{de2018blockchain}. Some studies on Decentralized Finance(Defi) pointed out that so-called decentralized finance is often not fully decentralized~\citep{ao2023decentralized, zhang2022blockchain}.

\subsubsection{Summary}
\ 
\newline
\indent
Some scholars divide the ethics of blockchain into three levels: blockchain technology design (micro-level), blockchain application (meso-level), and blockchain social impact (macro-level)~\citep{tang2019ethics}. Above, we briefly introduced the decentralization issues at the meso and macro levels. Such research on decentralization issues is very necessary because it is not only about what standards the underlying developers make to improve the code but also about people's confidence in decentralized concepts like blockchain and NFTs. Existing research is often purely speculative, and does not consider the perspective of users, which may make research fall into pure theory and ignore human perceptual factors. The participants in the "decentralized autonomy" pursued by blockchain supporters are the majority of blockchain community users. This is what our work tries to overcome.

\section{Methodology}
To obtain a holistic understanding of NFT stakeholders' perceptions and concerns about the NFT space, we applied a mixed-methods approach, leveraging both quantitative and qualitative inspections. 

Given the exploratory nature of our RQs and the limited literature, we first conducted a Twitter analysis on NFT-related tweets to provide a broad snapshot of the challenges, concerns, and opinions expressed by NFT stakeholders. Both Twitter analysis (see Section 3.1) and in-depth qualitative analysis (see Section 3.2) of the Twitter data were implemented at this stage. However, the findings of the Twitter analysis weren't informative in explaining the reasons behind the opinions of stakeholders due to the word limitation of tweets. To gauge stakeholders' lived experiences of identified concerns and potential solutions, we conducted semi-structured interviews (see Section 3.3). Combining the strengths of both quantitative and qualitative methods allows us to triangulate our results~\cite{tashakkori2007new}, enhancing the validity and reliability of the findings. Below, we elaborate on each part of the analysis.
\textbf{Data and Code Availability Statement}: We have made the dataset publicly accessible on Harvard Dataverse~\cite{DVN/GLFUPA_2023} and the accompanying code is available on GitHub at https://github.com/PlevanTem/NFT-cscw-2023 for open access.
\subsection{Twitter Data Analysis}
\label{sec:quan}
In this section, we mainly introduce how to obtain and analyze Twitter data. 
\subsubsection{Data Collection}
We chose Twitter as our site for data collection because NFT profile pictures have become symbolic social icons and trends on social media platforms~\cite{casale2022impact}, such as Clubhouse\footnote{Clubhouse: a voice-based social network.} and Twitter. These social platforms serve as primary arenas for public discussions and social activities related to the NFT industry, while Twitter data is the most accessible to researchers. By analyzing Twitter data, we are able to shed light on the perceptions and concerns of the stakeholders in the NFT market.

To collect the tweets, we used a scraper tool named SNScrape\footnote{\url{https://github.com/JustAnotherArchivist/snscrape}}. We queried tweets using "NFT" with five relevant keyword lists, regarding different concerns.
A case study on CryptoPunks\footnote{https://cryptopunks.app/} revealed a disparity in the average selling price between different genders and races, leading to a concern about equity, fairness, and diversity~\cite{sunmechanics,zhang2022visualizing}. Plus, a critical ethical analysis of NFTs brings our attention to harms, the environment, and legal and regulatory safeguards ~\cite{flick2022critical}. Moreover, the challenges of multi-factor (institutional, social, technological) trust in Bitcoin technology~\cite{sas2015exploring} also applied to the case of NFT assets due to their speculation attribute and decentralized technological service. Thus, we have lists of [fairness, equity, equality], [bias, discrimination, racism], [ethics, ethical, morality, moral], [trust, transparency, transparent], and [diversity, diverse].

Our search period ranged from 2021-06-01 to 2022-06-01. We excluded all retweets and only included tweets that satisfied the following criteria: (1) The tweets were English considering the computational cost of analyzing a large volume of tweets. (2) The tweets received at least one like to cut down on the huge amount of trivial posts. In the end, we collected 119,666 tweets along with their metadata such as retweetCount, likeCount, and followerCount. After preprocessing duplicated values and outliers, 112,308 tweets were identified and they were published by 69,543 accounts. 

\subsubsection{Data Analysis}
We began our analysis by conducting a word frequency analysis to identify core themes in public discussions. Frequently mentioned themes included "trust", "diversity", and "transparency" as seen in Section \ref{sec:topwords}. However, this approach provides the frequency of topics only. Thus, we conducted a qualitative, thematic analysis of the tweets by both Key Opinion Leaders (KOLs) and ordinary users to identify more practical insights toward answering RQ1.

\textbf{Top Words and Bigrams}
For the tweets scrawled, we intended to analyze latent themes and understand people's general concerns and perceptions towards the NFT market. We first pre-processed the tweets by the following steps: (1) removing hashtags, mentions, punctuations, URLs, and abbreviations; (2) transforming emojis to text; (3) replacing contractions with their full forms --- to achieve this, we manually created a common contraction-expression dictionary, e.g., "can't" and "can not"," "kinda" and "kind of"; (4) abandoning English stopwords. 

After lemmatizing and tokenizing, top words according to occurrence frequency were listed from high to low, which could provide us with a general understanding of trending topics. 

Bigram is a pair of co-occurrence words from a string of tokens, which is helpful to detect lexical trends~\cite{chaudhari2010lexical}. We plotted the top 20 bigrams for tweets retrieved with each keyword group and compared their connections and differences. The detailed bigram frequency lists are given in Table~\ref{tab:biagram-table}. The keywords shown in the bigram lists were inspirational for us to generate our interview questions. To understand the relationship among the top keywords, we used the python package Networkx\footnote{https://networkx.org/} to visualize the bigram complex network. 

\textbf{Identifying Opinion Leaders}
Opinion leaders are influential users whose opinions and ideas are widely discussed and shared on social networks and can exert a significant influence on other connected users\cite{PARAU2017157}. Given the significant influence of leading NFT communities and KOLs on the valuation of digital assets~\cite{casale2022impact,kapoor2022tweetboost}, and considering the frequent occurrence of hype, often misleading information on Twitter, we intended to examine the differences between ordinary users and opinion leaders regarding their social behaviors and opinions. 

In this regard, according to previous studies~\cite{semenkovich2019algorithms,PLOTTI2K12012opinionleaders}, parameters for inspection included the number of times an account is mentioned, the number of followers, the number of retweets received, and the number of mentions, which are the most relevant factors for detecting an opinion leader.  
Therefore, similar to a Twitter study ~\cite{Alexandre2021MakeTG}, we first identified a top 1\% subset of 173 accounts with followersCount, likeCount, and retweetCount more than 186076, 285, and 76 respectively so to cut out a large amount of data. We then created a network based on this subset. Each account is treated as a node and nodes are connected based on mentioning. 
We used Gephi\footnote{https://gephi.org/} version 0.9.7, an open-sourced graph visualization platform, to visualize connections in the opinion leader network. Centrality measures of nodes are used to describe the importance of KOLs. 

Randomly sampled tweets by different stakeholders, i.e., KOLs and ordinary accounts, are ready for the thematic analysis to compare the differences in their perceptions and concerns.

\textbf{Topic Modeling and Thematic Analysis of Tweets(Qualitative Twitter Analysis)
)}
To further address RQ1, we conducted a Latent Dirichlet Allocation (LDA) topic modeling on corpus extracted from tweets of KOLs and ordinary users. LDA is a generative probabilistic model for topic modeling, a popular way to extract common themes from large datasets~\cite{steyvers2007probabilistic}. This approach structures the data into three levels—word, topic, and document. LDA processes the input documents to yield topics as output. A topic is represented as a weighted list of words, representing common themes in a document. As part of our LDA model evaluation, we employed the $C_v$ coherence score from the Gensim package\footnote{https://radimrehurek.com/gensim/} to evaluate the LDA model and pick the hyperparameters that give the best performance, with 15 topics (KOLs) and 25 topics(ordinary users) identified respectively. This step facilitated the comparison of similarities and differences between perceptions and concerns from the lens of KOLs and ordinary users. However, considering topic modeling results lacked context of discussion. Thus, to have an in-depth understanding on the details of Twitter discussion, we conducted an additional thematic analysis by randomly sampling five tweets from each keyword in every topic, resulting in 750 tweets from KOLs and 1250 tweets from ordinary users. Following this, at least two authors independently coded the tweets, and then compared their coding results. If a discrepancy arises, a third coder would join to discuss the results and facilitate a final decision. This process was iterative and continued until consensus was achieved for all codes among coders no new codes were generated. In the end, we encoded 250 tweets from KOLs and 200 tweets from regular users. We consolidated the mutually agreed codes to formulate a codebook, where we ranked the top tweet codes from KOLs and ordinary users as indicators for their concerned topics.
\subsection{Interviews}
As mentioned earlier, discussions on Twitter often do not involve deeper discussions of the centralization-decentralization dilemma. Therefore, We recruited and interviewed participants from different backgrounds to get an in-depth understanding of NFT users' perceptions of the concerns and inform potential solutions.
\subsubsection{Recruitment}
Our recruitment targeted two main stakeholders of the NFT space, i.e., NFT creators and NFT collectors~\cite{sharma2022s}. NFT creators are people who create NFTs. NFT collectors are people who have experience collecting or trading NFT. Considering that NFT transactions are all conducted online, and the community is almost entirely online, we mainly look for participants through online recruitment. Additionally, we have an NFT collector on our team, through whom we also used snowball sampling to find participants. In conclusion, we used a combination of the following methods:
(1) Posting recruitment messages on Twitter, Discord, Wechat, and Weibo channels;
(2) Sending direct messages to NFT creators who promoted their NFTs on Twitter, Discord, Wechat, and Weibo; 
and (3) Snowball sampling.

In the end, we recruited 15 NFT users. 
Among them, 9 participants were recruited from Discord, WeChat, and Twitter channels, 4 from direct messages, and 2 from our personal contacts. They had different backgrounds, ranging from students and artists to engineers and financial practitioners. We believe these samples cover most of the general NFT market participants. More demographic information can be seen in Table~\ref{demographic}. 
\begin{table}[!htbp]
\begin{tabular}{lllllll}
\hline
ID  & Gender & Age   & Main Role & Year(s) of exp. & Profession        & Art Category \\ \hline
P1  & M      & 32    & Collector & 2 years        & Software Engineer & PFP/Fine Art \\
P2  & M      & 19    & Collector & 0.5 years      & Student           & PFP          \\
P3  & F      & 30-35 & Collector & 1 year         & Art Trader        & Fine Art     \\
P4  & F      & 30-40 & Collector & 0.5 years      & Business Employee  & N/A          \\
P5  & M      & 22    & Creator   & 1 year         & Student           & PFP          \\
P6  & F      & 30+   & Creator   & 1 year         & Artist            & PFP          \\
P7  & M      & 26    & Collector & 1.5 years      & Student           & PFP/Fine Art \\
P8  & F      & 25    & Creator   & 1.5 years      & Student              & Fine Art     \\
P9  & F      & 26    & Creator   & 1.5 years      & Artist            & N/A          \\
P10 & M      & 23    & Collector & 0.5 years      & Student           & N/A    \\
P11 & M       & 20    &Collector & 1 year    &Student          &PFP \\
P12 & M        & 21    & Collector & 0.5 years      &Financial Assistant    &PFP \\
P13 & M        & 24		& Collector & 0.5 years     & Student       &PFP   \\ 
P14 & F        & 25		& Collector & 1 year     & Software Engineer       &PFP   \\ 
P15 & M        & 19		& Creator & 1 year     & Artist       &N/A   \\  \hline
\end{tabular}
\caption{Demographic information of our participants. PFP: Profile Picture Project, N/A: no specific interest.}~\label{demographic}
\end{table}

\subsubsection{Interviews}
Participants were compensated for their time at 20 dollars per hour. Interviews were held online through Zoom or Tencent Meeting and lasted around 1 hour. The Chinese participants were interviewed by two or three Chinese-speaking researchers, and the interviews were conducted in Chinese. We recorded the interviews and saved transcripts of the conversations, which are affordances of Zoom and Tencent Meetings, upon participants' permission.  Participation in our study was completely voluntary, and participants could exit at any time. The research was IRB-approved.

The interviews were semi-structured. 
We started by asking about the basic background of participants and their interest NFT category. In order to understand their perceptions on ethical issues and centralized-decentralized dilemma, which is a frequent theme uttered by them, we asked questions about NFT trading, hacking and privacy, fairness and transparency, financial security, Ponzi schemes, etc. We also asked NFT creators specific questions about NFT creation.  


\subsubsection{Interview Analysis}
\label{sec:thematic_method}

After each interview, we held a brief meeting to discuss emerging themes. The iterative coding process was the same as that for qualitative Twitter analysis.
\section{Results}
The results explored the perceptions and concerns of stakeholders (RQ1), as expressed on social media, through both quantitative and qualitative insights derived from the Twitter analysis. Additionally, a subsequent section featuring interviews with 15 participants, including NFT collectors and creators, not only unveiled more in-depth information about their perceptions and concerns that were initially identified in the Twitter analysis but also broadened the scope of interview questions regarding potential solutions to answer RQ2.

\subsection{Quantitative Findings from Twitter Analysis}
\subsubsection{Top Words and bigrams}
\label{sec:topwords}
\begin{figure}[H]
     \centering    
     \includegraphics[width=0.5\textwidth]{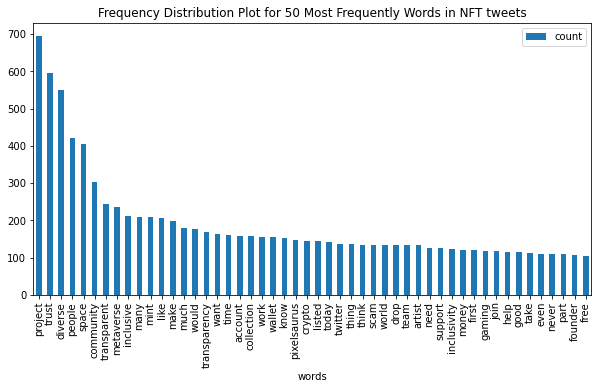}
    \caption{The most frequent 50 words in NFT ethics related tweets; B: Top keyword network of NFT Trust-group.}
     \label{fig:topwords_network}
 \end{figure}

There were 7106 samples and 46935 outcomes in the processed word list from collected tweets. Based on the occurrence of words shown in Figure \ref{fig:topwords_network}, we identified three categories of public concerns that were most frequently mentioned in the NFT market, namely, Trust, Diversity, and Transparency.

\textbf{Trust and Transparency} \textbf{Trust} is a crux problem facing the current NFT market. It was mentioned by the second highest, much more frequently than other keywords. Notably, transparency is an integral part of building stakeholders' trust in the NFT marketplace. The bigram analysis for the "trust" category (Table~\ref{tab:biagram-table}) reveals that openness and transparency about an NFT project's team and roadmap are key factors in fostering trust and avoiding instances of scams.
In decentralized and emerging markets like the NFT market, it is common and crucial for NFT projects to provide clear and accurate information to users on social media and their websites. Such transparency indicates their reliability and contributes to a more trustworthy environment for potential NFT users. 

\textbf{Diversity, Inclusivity, and Equity.} The bigram analysis for the "diverse" category (Table~\ref{tab:biagram-table}) illustrates the public's high expectations for a diverse and inclusive NFT community. The most frequently mentioned words, such as "love," "women," and "love", indicate that the community values inclusivity, diversity, positivity, and unity, as opposed to the rigid and centralized structure of the traditional art market \cite{baytas_stakeholders_2022}. However, the frequent use of such buzzwords as "diverse" and "inclusive" in the NFT community could be seen as a form of propaganda, as later thematic analysis revealed \ref{ta_nft_promotion}.

\subsubsection{NFT Opinion Leaders on Twitter}
\label{sec:kol_network}
\begin{figure}[H]
    \centering
    \includegraphics[width=0.5\textwidth]{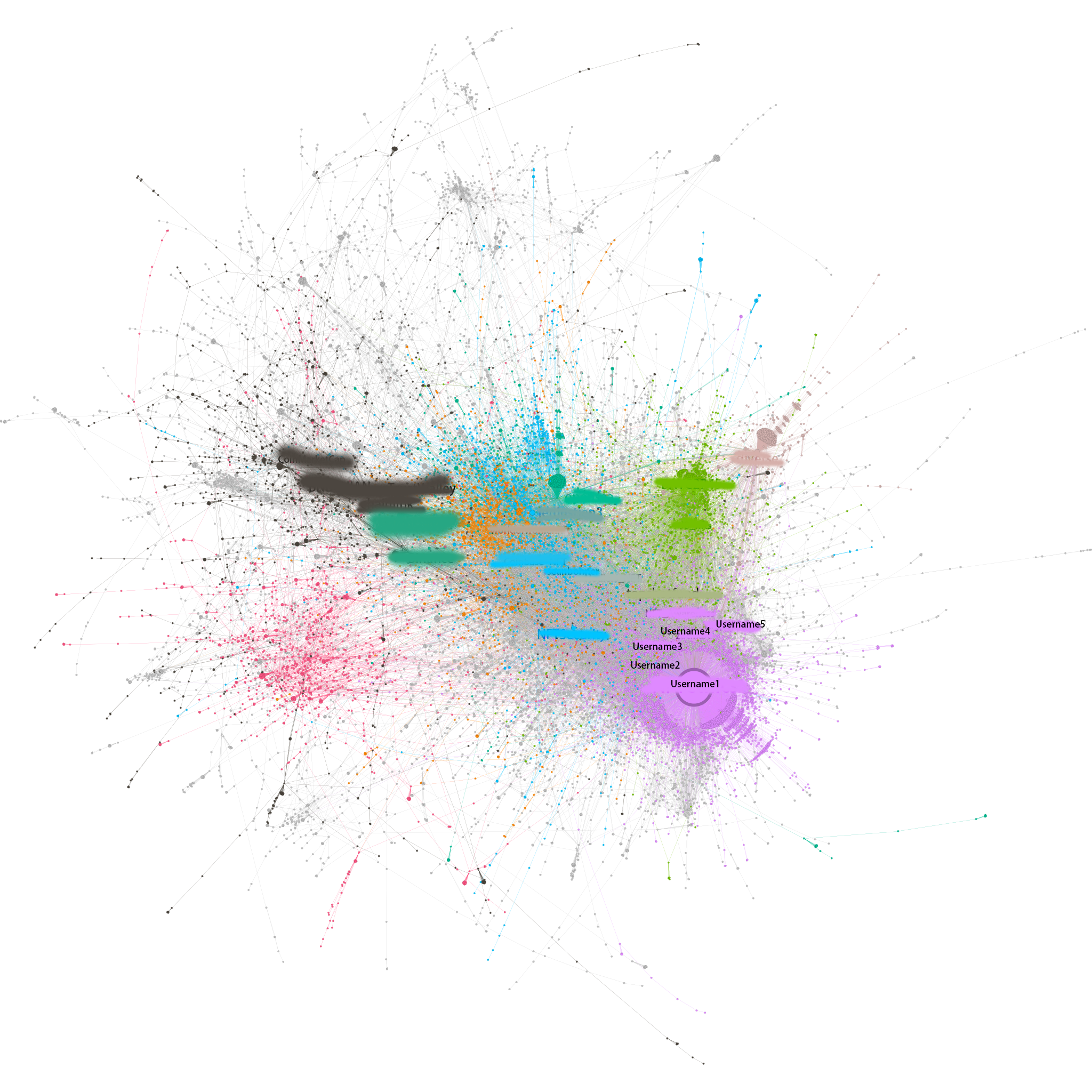}
     \caption{Network Visualization:  Opinion Leaders in NFT Marketplace. Label size reflects eigenvector centrality. Node size reflects follower count. Node colors represent detected communities.}
     \label{fig:kol_network}
\end{figure}

Through the network analysis, we identified 173 influential KOLs that often spoke out opinions and concerns about the NFT marketplace. The top 20 KOLs are shown in Table~\ref{tab:top20nodes}. These key individuals span a range of roles within the NFT community, including project owners 
and NFT 'vigilantes' (see explanation in Section \ref{sec:kol_tweets}) 
, as well as collectors 
. They also affiliate with various projects 
on platforms like Opensea, Binance, and TheSandboxGame. Positioned at the heart of the KOL network (see Figure \ref{fig:kol_network}), @Username1\footnote{The Twitter handler names mentioned in this paper have been replaced with pseudonyms to protect the privacy and anonymity of the individuals involved.} imposed a larger influence and enjoyed a higher follower count compared to other users. By differentiating between KOLs and ordinary users, we were able to identify and compare the primary discussion themes between these two groups of NFT stakeholders (see the comparison in Section \ref{sec:qual_TA}).

\subsubsection{Topic Modeling}
\label{sec:topic modeling}
Our analysis of the LDA topic modeling for both KOLs and ordinary users' tweets reveals distinctive perceptions and concerns within each group. See their topics in Appendix \ref{app:topic_modeling}. KOLs predominantly discuss topics related to the potential of NFTs in a more diverse and inclusive space(topics 0-14), the trust and transparency(topics 0-14) necessary for these projects, and metaverse(topics 1, 2, 3, 5, 9, 11, 12, 13) - a virtual reality space where users can interact with a computer-generated environment and other users. The recurrent theme of the metaverse indicates KOLs' strong interest in the intersection of NFTs and the development of virtual worlds, implying a priority focus on the implications of NFTs for the future of digital environments, such as to prove ownership of digital assets in the metaverse. In contrast, ordinary users express concerns and interest in a broader range of subjects. They discussed the trustworthiness and transparency of NFT projects, tangible challenges in transacting NFTs, and the tangible role of NFTs in digitized art and gaming.

A cross-examination of these discourses shows a shared concern about the trustworthiness and transparency of NFT initiatives across both groups, revealing a universal acceptance of these attributes as the bedrock for NFT operations. This shared concern signals a need for secure, reliable, and clear transactions and operations within the NFT market. This could also reflect broader debates surrounding decentralization, as trust and transparency are cornerstones of decentralized systems. However, noticeable differences also emerge. KOLs seem more preoccupied with the broader implications of NFTs in creating diverse, inclusive, and ethereal digital spaces, while ordinary users frequently grapple with the practical aspects and immediate challenges of NFTs, such as issues with wallets(topic 5, 7, 14, 28), buying and selling NFTs(topic 1, 3, 7, 11, 23, 24), and their role in the contemporary art and gaming industries(15,16,19).

\subsection{Qualitative Findings from Twitter Analysis}
\label{sec:qual_TA}
 \par
The above quantitative analysis of tweets gave us an overview of the commonly expressed concerns of the NFT market. 
An in-depth understanding of the opinions of both KOLs and ordinary users was presented in the following thematic analysis part, which complemented the results of topic modeling. We identified 31 themes from the KOLs' tweets and 72 themes from the ordinary users' tweets. See the thematic coding in Appendix \ref{codebook}, with the top 20 coding topics shown in figures \ref{app:topics_KOL} and \ref{app:topics_Od}. The results aligned with our quantitative results --- both KOLs and ordinary users frequently mentioned transparency-, trust-,  and diversity-related topics. Similar to topic modeling \ref{sec:topic modeling}, tweets from KOLs showed more concern on high-level issues such as transparency, discussion about unethical behaviors of projects (e.g.. scam, rug pull, fishing, etc.), trust, community etc., while ordinary users presented a more focus on financial topics such as promotion and investment (see figures \ref{fig:kol_thematic_analysis} and \ref{fig:od_thematic_analysis}). By reading the randomized tweets, we identified some patterns between the roles of KOLs and ordinary users in their Twitter conversations.

\subsubsection{Tweets from KOLs.}
\label{sec:kol_tweets}

\begin{figure}[H]
    \centering
    \includegraphics[width=0.8\textwidth]{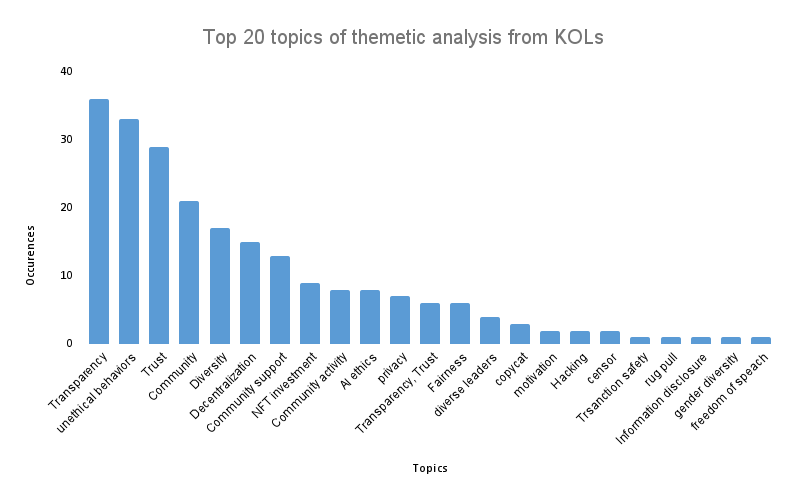}
    \caption{Top 20 topics of thematic analysis from KOLs}
    \label{fig:kol_thematic_analysis}
\end{figure}

\textbf{KOLs as Vigilantes: Promoting Transparency and Safety in NFT Transactions.}
Transparency, trust, unethical behaviors, and community supervision are concerned for around 53\% of the thematic coding results from KOLs' tweets (see Figure~\ref{fig:kol_thematic_analysis}). Do Your Own Research (DYOR) \footnote{https://academy.binance.com/en/glossary/do-your-own-research} is a common crypto slang because stakeholders normally believe that investors should conduct extensive research before investing due to the fast and easy spreading of misinformation. Therefore, influencers can play an important role in NFT community supervision, providing warnings for risks such as suspicious scam projects, and promoting transaction safety, and financial behaviors(e.g. rug pull, fishing, scam, hacking, etc.) from stakeholders. We regard this type of ethic investigator as \textit{NFT vigilantes}. We can see from the KOL network \ref{sec:kol_network} that nodes surrounding @Username1 including @Username2,@Username3,@Username4, and @Username5, etc. formed a vigilante cluster, playing a critical role in preserving safety and promoting transparency within the NFT landscape by revealing misinformation. The node @Username1claimed to \textit{Uncover inconvenient truths about this digital kingdom and its most influential players in order to address the ethical issues and foundations of Web3}. For example, @Username1warned the NFT community about emerging scam projects and attacks targeting people's funds in the following tweets:
\begin{quote}
    \textit{"Disclaimer: we have already shown today that lots of `scam projects' are using similar kinds of 3D images for the Metaverse (cats, bears, pigs). The above picture shows that it is not only limited to those projects but to a much wider variety of projects that copy legitimate ones."}
    \newline
    \textit{"We have noticed lots of new people joining this space that are very naive. It is not uncommon for the funds to be distributed to non-anon [non-anonymous] accounts or accounts with a history."}
\end{quote}

\begin{figure}[H]
\centering\includegraphics[width=0.8\textwidth]{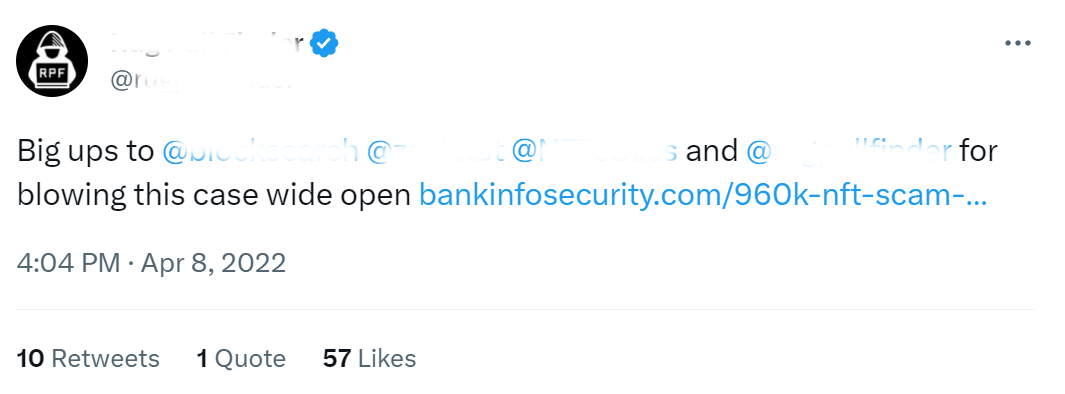}
     \caption{KOLs work together to showcase scamming projects.}
     \label{fig:kol-showcase}
\end{figure}

\begin{figure}[H]
\centering\includegraphics[width=14cm]{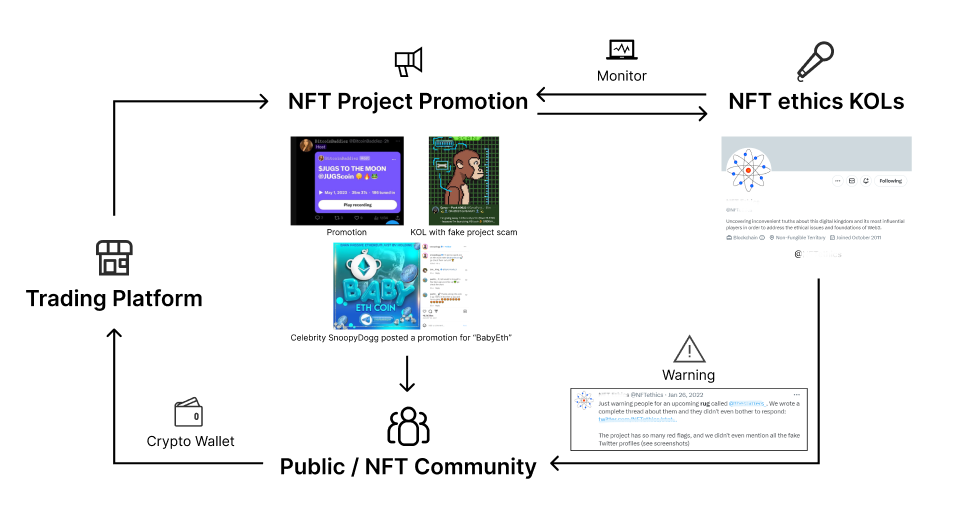}
     \caption{Decentralized Gatekeeping: The Role of Influential NFT Community 'Vigilantes' KOLs in Detecting Risk and Promoting Transparency on Twitter.}
     \label{fig:user-story}
\end{figure}

See another example in Figure \ref{fig:kol-showcase}, KOLs @Username6, @Username3, @Username7 @Username4 worked together to reveal a French social media Influencer behind an NFT rug pull \footnote{https://www.databreachtoday.asia/960k-nft-scam-affects-nearly-1200-victims-a-18853} which caused \$960K NFT scam affects nearly 1,200 victims. In summary, we illustrated the activity pattern of this vigilante type of KOLs in the NFT community on Twitter as shown in Figure \ref{fig:user-story}. In this decentralized and less regulated NFT market, they usually practice as a decentralized source of credit information and sleuth surveillance on NFT projects and teams to promote Web3 spirits such as transparency.

\begin{figure}[htbp]
    \centering
    \begin{minipage}{0.6\textwidth}
        \centering
        \includegraphics[width=1\textwidth]{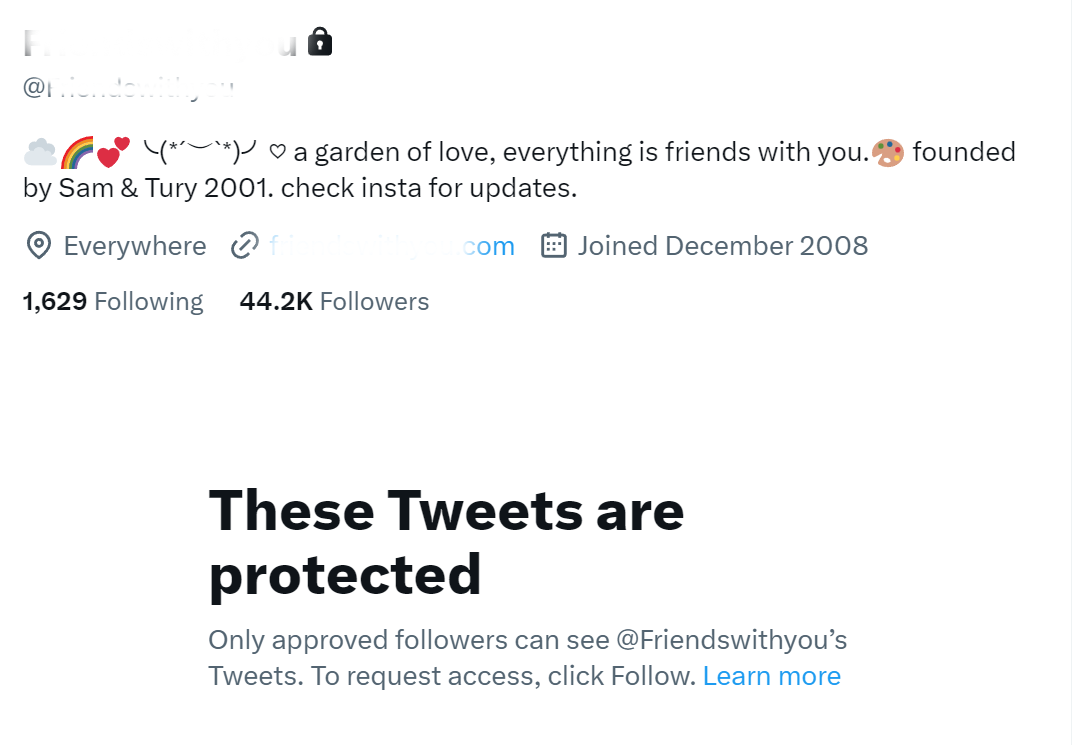} 
    \end{minipage}\hfill
    \begin{minipage}{0.3\textwidth}
        \centering
        \includegraphics[width=1\textwidth]{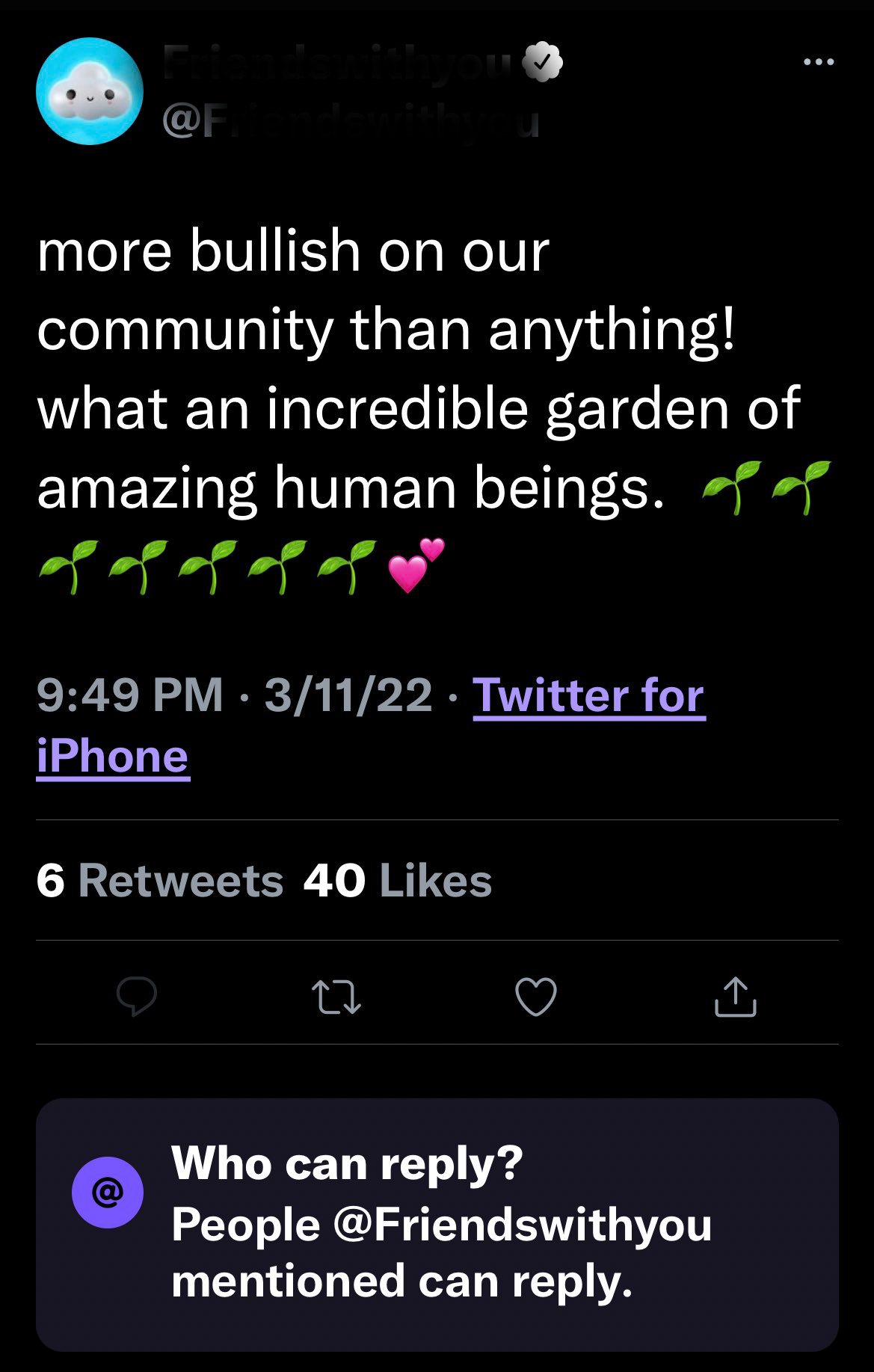}
    \end{minipage}
    \caption{Example of an NFT project's violation of freedom of speech. Left: only approved followers can see and express their opinions. Right: only specific people can reply.}
    \label{fig:violation}
\end{figure}

\textbf{The Paradox: Advocacy vs. Reality in the NFT Community} The early NFT space is founded on the visions of decentralization. The NFT community tended to promote the perks and superiority of their decentralized organization in the name of being inclusive and democratic. However, the reality is not as blissful as what the Web3 advocates expected. Not all NFT stakeholders adhered to the professed principles. Actions such as blocking and censoring are considered as betraying Web3 ethos, however, were done by some NFT projects such as Figure \ref{fig:violation} shown. 

This paradox also existed among KOLs. As one tweet in our thematic codebook showed, the leading KOL @Username1 criticized some big NFT influencers for violating the spirits of Web3 by blocking or muting them on Twitter. 
\begin{quote}
     \textit{@Username1 posted "Just a quick reminder that we have never blocked/muted anyone on Twitter and never will. It's one of the fundamental characteristics of Web3. Many other big influencers have blocked us, all pretending to be Web3 evangelists, but they don't practice what they preach."}
\end{quote}
Sarcastically, some comments under this post attacked @Username1, accusing the KOL of engagement farming\footnote{Engagement farming is when one wants to get more engagement on social media platforms.}, while some Twitter users supported @Username1. This reveals a discrepancy between theory and practice, with some users supporting @Username1, and others criticizing them. 
\begin{quote}
     \textit{A user criticised @Username1by saying, "Just seems like throwing mud for no reason. You have legit claims against these accs or are you just salty not everyone likes what this acc has turned into?"}

     \textit{Another vented, "When the fuck did 'being forced to listen to everyone's shitty takes' become a pinnacle of Web3."}
     
     \textit{Another reply supported @Username7 mocked the situation, "Noticed that people commenting on this tweet were the same people praising you guys when you exposed beanie. However, when the person involve is part of their circle, they suddenly became defensive and called you engagement farmer."}
\end{quote}
These divergent responses expose a wider paradox within the NFT community – while it promotes ideals of decentralization and open dialogue, its actions often contradict these principles, leading to discord and conflicting attitudes.

\subsubsection{Tweets from Ordinary Users.}
\label{sec:od_tweets}
\label{ta_nft_promotion}
Ordinary NFT users, compared to KOLs, covered a broader range of topics regarding their specific concerns, including phishing, environmental impact, art privacy, etc. However, touting NFT projects and posting advertorial tweets to promote NFTs are commonly seen. Note that such tweets out for lucrative purposes occurred in both ordinary users and KOL groups.

\begin{figure}[H]
    \centering
    \includegraphics[width=0.8\textwidth]{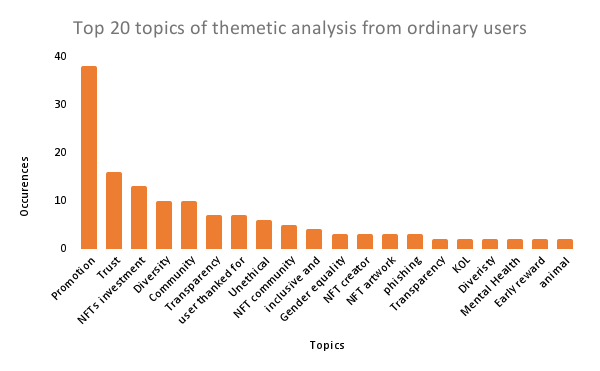}
    \caption{Top 20 topics of thematic analysis from ordinary users}
    \label{fig:od_thematic_analysis}
\end{figure}

\textbf{Financial Speculation in NFTs.} Many people entered the NFT space purely for financial reasons. Ordinary users cared about NFT investment, profit, newest drop-in\footnote{The release of a new NFT.} more than ethical issues. We can tell from Figure \ref{fig:od_thematic_analysis} that ordinary users discussed twice as much about NFT investment as ethical concerns. This opportunistic situation blended with new lucrative forms of technology has led to a hotbed for unethical behaviors, deeply concerning stakeholders in this industry. As the Twitter account @Username8 commented,
   \begin{quote}
       \textit{"Unfortunately, I've seen honest polls on nft Twitter where people are clearly here to make money,  even at the expense of others and regardless of the morality behind that profit."}
   \end{quote}
  
\textbf{Artists in the NFT Space: The Appeal of Community and Shared Ownership.} Apart from profits,
another reason for artists being drawn to the NFT space, based on tweets, was the community aspect, which differentiated the NFT market from traditional art markets. The NFT community provided both artists and collectors with a sense of belonging and a bridge to communicate. According to an NFT artist, @Username9,
     \begin{quote}
         \textit{"I Jumped into web3/nfts to focus on building, connecting and moving my art/music/photography away from what i felt was a bias\&rigged system."}
    \end{quote}
        
 A similar opinion was expressed by another user,
\begin{quote}
    \textit{"community
    - a group of people living together and practicing common ownership. This move has never been done and can place this project in art history, that's the key in the NFT space this early."}
\end{quote}

\textbf{Leveraging Trust and Diversity: The Marketing Strategy of NFT Projects.} 
Vision and unique value proposition of NFTs are marketing tools of NFT project owners. About 38\% of the coded tweets by ordinary accounts were about NFT promotion. 
Instead of focusing on the authenticity of NFT artworks, public trust is believed to be a critical factor for a sustainable NFT project. Transparency, diversity, inclusion, and gender equality are also frequently mentioned values according to Figure~\ref{fig:od_thematic_analysis}. According to @Username9, the community manager of NFT Immutable X, 
  \begin{quote}
     \textit{"Price, market sentiment \& macro factors don't scare the best projects... Place trust in teams who focus on the bigger picture, they are the ones who are truly GMI."}
  \end{quote} 

Apart from attracting users with values and visions, the reputation of the team behind an NFT project is also an important attraction, as @Username10 said, 
  \begin{quote}
     \textit{"As an NFT collector and artist, I’ve learned that it's not only about investing in the NFTs themselves, but investing in the good people behind the projects."}
  \end{quote}

\subsection{Qualitative Findings from Interviews}
\label{sec:interview}
Through sections 4.1 and 4.2, we have a preliminary understanding of the issues in the NFT market that stakeholders are concerned about. In this section, we first further summarized and revealed the participants' concerns about the NFT market. For example, trust and transparency are often closely related to financial risks, while diversity is often related to ethics and discrimination. We grouped these issues into four categories: financial risks, counterfeit NFTs, hacking, and traditional unethical NFTs.  We then exhibit participants' perceptions of centralized regulations in different issues.  

\subsubsection{Issues in the NFT Market of Participants' Concern}
\ 
\newline
During the interviews, we found that participants had different concerns about issues in the NFT market. Almost all participants, especially collectors, were concerned about financial risks. P1 sharply pointed out:
\begin{quote}
    \textit{"In fact, no one who trades in the NFT market dares to say that he is not here to make money, and everyone expects that he will not be the greatest fool."}
\end{quote}
This suggested that the NFT market is currently flooded with a lot of speculation. 

All creators were concerned about the counterfeit NFTs, and they believed that their related rights had been compromised. However, collectors, were more concerned about whether NFTs could bring them financial benefits than whether they purchased counterfeit NFTs. In section 4.3.2, we can see the differences in ideas between the two parties (i.e., KOLs and normal users) caused by their different interests. 

Two participants (P3, P6) had their Ethereum wallets stolen, and they admitted that such incidents have greatly affected their confidence in the NFT market. Most other participants had no experience of being hacked. P5 further pointed out that hackers were only a small part of the NFT market, and it would not play a decisive role in people's confidence in the NFT market.

Although we used ethical keywords to collect related tweets, most of the interview participants (12/15) had only heard a little about traditional ethical issues in the NFT market, such as racism, sexism, pornography, violence, bias, etc. Three participants (P3, P6, P8) had a deep understanding of unethical NFTs. P3 and P6 were more disgusted about pornographic tolerance. P6 mentioned that in addition to discrimination and bias, there may also be NFTs about pornography, guns, and violence. However they admitted that such inappropriate content also existed in traditional digital artworks, and the ethical issues in NFTs were not special. P8 mentioned that there were fewer female artists in the NFT market, and admitted that this issue could not be said as a unique issue in the NFT or Web3 field, but instead a common gender issue in the whole society. 

To summarize, participants paid the most attention to financial risks, and the least attention to traditional ethical issues in NFTs. During our interviews, we also found that most of the participants (12/15) did not understand the blockchain principles behind NFTs, which suggested that software developers only accounted for a small part of the NFT community.

\subsubsection{Participants' Views on Centralized Regulation in the NFT market}
\ 
\newline
We found it difficult to obtain users' views on the centralized regulation and decentralized autonomy of the NFT market by collecting Twitter data. The discussion on decentralization on Twitter was superficial, but it was actually the biggest difference between NFT and traditional e-commerce platforms. In this section, we report participants' views on the need for centralized regulation to address the four main issues in the NFT market. The participants had diverse perspectives on different issues.

\textbf{Centralized regulation for preventing traditional unethical NFTs.}
\label{sec:4.3.1}
As mentioned above, Opensea is a centralized platform that has the authority to remove NFT works from the front end. So is this centralized management necessary? Has it violated the right of some creators to freely propagate their works? Participants were deeply divided on this issue. Four participants supported the existence of a lower level of centralized regulation, and eight other participants opposed it. Opponents of centralized regulation argued that unethical NFT propagation should be prevented through community autonomy instead of centralized regulation. P15 acknowledged that when seeing disgusting NFTs or those he did not like, he still did not think it was his right to take them down: 
\begin{quote}
    "If somebody doesn't like something, then just don't watch. You're not being forced to consume it or interact with it. So I would say, nobody has the right to take any piece of work down. Even if it's disrespectful, even if it's illegal, people have the right to do whatever they want with their own personal lives."
\end{quote}

P8, P10, and P14 thought unethical NFTs were difficult to define, so no accurate conclusions could be drawn. P8 said:
\begin{quote}
    {"This kind of ethical conflict and controversy is an inevitable result of decentralization. I even said that I hoped they have no right to publish, but I have to say where is the boundary of this right? When you mention the Nazis in Germany and the Nazis in China, the destructive power it produces, or the negative impact it produces, is actually not in the same order of magnitude. So in different cultural contexts, what is not allowed to talk about, what is allowed to talk about, this is a very delicate thing."}
\end{quote}
\ 
\indent
P7 pointed out that centralized regulation was necessary in the current stage, but can be gradually relaxed over time. He thought that we were now in a transitional stage from centralization to decentralization. With the development of technology and the improvement of the form of autonomy, the decentralized regulation of unethical NFTs will gradually become possible.
\ 
\newline
\indent
\textbf{Centralized regulation for reducing financial scams.}
As mentioned before, Twitter users most cared about transparency and trust. Some unethical NFT creators and KOLs often lured buyers into investing by spreading disinformation or misinformation through social media. Is it necessary to mitigate such misconduct from publishing information and recovering property through centralized regulation? P7 held the opinion that regulation was a must, and he said:
\begin{quote}
    {"Our world is not a world of the jungle. Supporting the need for no regulation is like a woman who wears very little meets a pervert, and is accused of wearing too little. This is victim guilt."}
\end{quote}
\ 
\indent
P1 and P2 were not opposed to government involvement, but said they did not want authorities to get involved too deeply, P1 said:
\begin{quote}
    {"After the influx of a large number of people, the code of conduct should be followed... government limited intervention will improve efficiency. It is best to be only targeted to the criminals."}
 \end{quote}
P10 believed that centralized regulation would undermine the freedom of speech, and had reservations about centralized regulation.
The remaining participants were opposed to centralized regulation, and their reasons were similar. First, they believed that public opinions in the community could effectively play a supervisory role. Second, centralized regulation would undermine freedom of speech and have worse consequences.
\ 
\newline
\indent
\textbf{Centralized regulation for preventing hacking.}
As mentioned above, some decentralization proponents argue that once code is deployed on-chain and accepted by a majority of miners, it cannot be changed. Whether it is the theft of private keys or the exploitation of smart contract vulnerabilities, it is part of decentralization. In sum, "Code is law~\citep{de2018blockchain}". 

Eight Participants supported "Code is law". P5 thought the expansion of centralized power would eventually lead to decentralization in name only and damage user rights. P3 mentioned that she was a "code believer" and believed that rules could be made through code. 

However, other participants still supported centralized regulation to prevent hacking. P8 thought it was difficult to achieve "code is law" at present, saying:
\begin{quote}
    \textit{"Ideally, I agree that code is law, but now the speed of development of fraudulent methods is far beyond the code, even if you use centralized regulation, you may not be able to get your property back."}
\end{quote}
\ 
\indent
P10 further pointed out the logical doubts of radical decentralization proponents: the code of smart contracts was originally written by people, but it now became a paradox that the code was required to override all people.
\ 
\newline
\indent
\textbf{Centralized regulation for preventing counterfeit NFTs.}
This issue is similar to traditional unethical NFTs. However, different stakeholders had very different perceptions on this issue. All creators, even the most radical ones (P5, P15) who supported decentralization, admitted that counterfeiting NFTs required centralized regulation. P5 said:
\begin{quote}
    \textit{"This is an issue that bothers me very much. I often think that although centralized regulation can prevent counterfeiting of NFTs, it goes against our decentralization thinking. I don't think there is any other better way to solve it so far."}
\end{quote}
P15 said:
\begin{quote}
    \textit{"I'd have to prove that it was made by me. First, I posted it online, either saved files or any photos. And second of all, I would immediately reach out to the platform. I would let them know: here's my thing, here's proof that I made it, and I have very distinctive stuff. If the platform is not responding, I would say I would reach out to the community. I would tag people. I would let the influencers know. I let the artist who stole my artwork know that I know. And most of the time it works."}
\end{quote}

For collectors, perceptions quite varied, but they did not support centralized regulation except P2. P1 and P4 thought that centralized regulation was too difficult. P7 and P10 thought that centralized regulation was not necessarily required, and AI models in the form of smart contracts could be deployed to achieve the purpose of identifying counterfeit NFTs. 

\textbf{Summary of perceptions.} We briefly organize the results of Section 4.3.2 to get Table \ref{Perceptions}. The first column in the table includes specific issues, and each subsequent column represents the number of people with an identity who support centralized regulation. 
\begin{table}[!htbp]
    \small 
    \begin{tabular}{@{}cccccc@{}} 
    \hline
    Unethical Behaviors  & Positive & Neutral & Negative & Positive Creators & Positive Collectors\\ \hline
    Financial Scams and Disinformation  &  3  &  1    & 11 & 0 & 3       \\
    Hacking  &  7    &  0    &  8  & 3  & 4      \\
    Confeiting NFTs  & 5    &  0    &  10  & 5  & 0        \\
    Transitional Unethical NFTs  & 4    & 3 & 8 & 1 & 3    \\
    \hline
    \end{tabular}
\caption{Perceptions towards centralized regulation of different issues.}~\label{Perceptions}
\end{table}

\rev{
\subsubsection{Thought Experiment on NFTs and Traditional E-commerce Platform}
\ 
\newline
Consider an example. After a musician creates a digital album, he can sell it as a commodity on Spotify (a digital music service platform), or he can mint it into an NFT and sell it on Opensea. Since it is difficult to find the same digital asset being sold on both traditional e-commerce platforms and NFT platforms in the market, we conducted a thought experiment in the interview. For collectors, we asked them if the same artwork existed, would they buy it on a traditional e-commerce platform or an NFT platform? We assume that the price of the same digital asset after being minted into NFT will be more expensive than in traditional e-commerce. This is because NFT minting requires additional power consumption and causes environmental problems~\citep{kapengut2022event, zhou}. For creators, we asked them why they chose to create NFTs instead of selling them on traditional e-commerce platforms. Most collectors are not willing to buy NFTs in this situation, or they only buy them for speculation. Only P3 chooses to make NFT purchases under non-speculative circumstances:
\begin{quote}
    \textit{"I would choose to buy NFTs. Because this is a new thing and is in line with the historical development process"}
\end{quote}
For creators, there are three reasons why they enter the NFT market: 1. Recognize the concept of decentralization and believe that traditional, centralized companies are not good for creators. Friendly (P5, P15, P9). For example, P5 once worked in an art and creative company. The company took most of his income and forced him to work overtime. 2. After being introduced by a friend, followed the trend of the times and joined (P6). 3. It is believed that the relationship between NFT creators and collectors is not that of a traditional merchant and customer, but more like a collector investing in a creator that he admires to establish a relationship with the creator (P8).
}

\subsection{Summary and Answering RQs}
In this section, we summarized answers to our two research questions
based on the results from both quantitative and quantitative methods.

\subsubsection{Perceptions and concerns of stakeholders in the NFT communities (RQ1)}
Through cross-validation using two different methods, we understand the perceptions and concerns of some NFT market participants. We found stakeholders had different perceptions of different issues.

Financial risk was the most concerning issue in both tweets analysis and interviews. Twitter users were most concerned about transparency, trust, and diversity. However, transparency and trust were often related to investment in this context -- most tweets emphasized transparency and trust in order to expose NFT projects with financial risks. Moreover, some tweets took advantage of the common values (such as feminism and Black Lives Matter) of the NFT market for advertising. In the interviews, most of the collectors admitted that they bought NFTs for investment purposes. So it could be inferred that most stakeholders emphasized transparency, trust, and diversity in order to buy NFTs with more room for appreciation or sell NFTs with a high price --- their ultimate goal was money. The main body of collectors in the market were speculators, making the entire market more like a Ponzi scheme and providing an environment for financial scams. This also explains why stakeholders were most concerned about financial risks.

In Twitter analysis, a small number of users also mentioned hacking in tweets. Hacking often co-occurs with financial scams. In the interviews, it can be found that compared with financial scams more participants (7/15) supported centralized regulation to address hacking. One possible reason was that hacking would bring clear losses to users, while financial scams could be easily confused with financial bubbles. Unlike scams, bubbles had no intention of misleading investment but were just a phenomenon caused by investors' excessive expectations~\citep{chalmers2022beyond}. This may also be the reason why more participants supported centralized regulation to address hacking rather than financial risks.

Counterfeit NFTs were another issue that required attention. In Twitter analysis, some counterfeit NFT-related content also appeared. In the interviews, creators have expressed strong concerns about counterfeit NFTs. However collectors often have a different attitude towards counterfeit NFTs, and they are indifferent to the centralized regulation of counterfeit NFTs and even deliberately buy counterfeit NFTs. This is likely due to the different interests of the two parties. Counterfeit NFTs seriously infringe the copyrights of creators, but for speculators, whether they are counterfeit works does not matter, and what matters is that the NFTs can be sold for a high price. Most collectors are not willing to buy NFTs in this situation, or they only buy them for speculation. Only P3 chooses to make NFT purchases under non-speculative circumstances:


Twitter analysis also exhibits a fair amount of diversity and feminism, but these contents are often just slogans to promote NFT projects. During the interviews, the participants also mentioned some traditionally unethical NFTs, but in general, stakeholders were not concerned about these issues.

\subsubsection{Is blockchain the panacea for solving those issues in the NFT market? (RQ2)}
We found little discussion on centralized regulation and decentralized autonomy in tweets. Therefore, stakeholders’ perceptions of centralized regulation and decentralized autonomy all come from interviews. Generally speaking, in the NFT market, most interview participants still insist on decentralized autonomy for existing issues. Hacking is the issue that most participants support centralized regulation, but only nearly half of the participants(7/15) support centralized regulation. Participants have slightly different perceptions on different issues. For financial scams, participants hope to solve this problem through community supervision rather than centralized regulation; when it comes to hacking, about half of the participants support centralized regulation; when it comes to counterfeit NFTs, all creators support centralized regulation, while collectors support decentralized autonomy; when it comes to traditional unethical NFTs, there are more neutral perceptions. This shows that participants have different altitudes of centralized regulation regarding different issues. On certain issues, people with different identities and interests may have completely opposite positions. For example, creators support the centralized regulation of counterfeit NFT, while collectors are relatively indifferent. The most typical examples are P5 and P15(they are creators): they are extreme supporters of decentralization, but once their personal interests are damaged, they still expect or at least use centralized regulation. 

Thus, we think that decentralization is not a panacea for all issues in the NFT communities. People with different interests often have diverse views on different issues. In a thought experiment, most collectors are not willing to pay for the concept of decentralization. we also learned that centralized regulation and decentralized autonomy had their own advantages and disadvantages from interviews. For specific stakeholders and specific issues, decentralization can even be harmful. This is what our research found: the centralization-decentralization dilemma. Its essence is the dilemma caused by the conflict between the slogan of decentralization and the interests of stakeholders.

The collective resistance of creators against counterfeit NFTs and even abandoning their support for decentralization are a good illustration of the centralization-decentralization dilemma faced by NFTs. From a broader perspective, NFT uses decentralized blockchain technology to distinguish it from traditional artworks and e-commerce platforms, but it has to adopt centralized regulation on some issues, which is also a centralization-decentralization dilemma. When facing complex social and ethical issues, decentralized technologies such as blockchain cannot be used as a general method to solve all issues once and for all. We will discuss the scope of the application of decentralized autonomy in the NFT market in detail in the discussion section, and give our thoughts on centralization-decentralization dilemma.

\section{Discussion}
Combining previous research, we propose potential solutions to resolving the concerns in the NFT market, Looking ahead to the future of NFT, and discuss the limitations of our research to shed light on future studies.

\subsection{The Direction of Solutions to The Concerns in NFT Market}
In this subsection, we combined Twitter data analysis and interview results to come up with our recommendations for resolving centralization-decentralization dilemma in the NFT market.
\ 
\newline
\indent
\subsubsection{Incorporating Centralized Regulation for Destructive Issues} 
\ 
\newline
\indent
Since the birth of the blockchain, its concept of decentralization is controversial. However, through blockchain's continuous development, more and more people have recognized the concept of decentralization. Whether centralization or decentralization, it is only a means, not an end. Decentralization is created to solve a series of drawbacks(For example, the bad debts of banks during the 2008 financial crisis) caused by centralization, and its purpose is to better guarantee the rights of decentralized users. However, the current underlying technology of blockchain is not enough to support large-scale decentralized autonomy~\citep{de2018blockchain}. Technical decentralization does not necessarily guarantee substantive decentralization~\citep{schneider2017once, atzori2015blockchain, dodd2018social, qin2021cefi, ao2023decentralized, zhang2022blockchain}. Events that benefit from blockchain technology are also not uncommon.~\citep{fu2022ai} If decentralization seriously damages the rights of users, the NFT market will have a crisis of trust. If it is only technically decentralized but essentially centralized, then this runs counter to the ``trust building'' that decentralization promotes.
 
Therefore, it is reasonable to apply limited centralized regulations. Issues caused by technological decentralization are often easily solved under centralized regulation. The subject of this regulation can be the government or other organizations. However, the authorities of these centralized agencies need to have clear boundaries. Nearly half of our interview participants support centralized regulation to address hacking, and all creators support centralized regulation to mitigate counterfeit NFTs. This shows that the existence of these two issues has indeed affected the trust of the NFT market, so it is reasonable to conduct centralized regulation on these two issues. \rev{Centralized regulation is especially needed for NFT hacking using phishing websites. We mentioned in related work that phishing websites deploy malicious code on the front-end websites to steal users’ private keys and property. This behavior is not essentially different from traditional hacking and it often happens outside the back-end blockchain. It is obviously wrong to confuse the centralized front-end websites with the decentralized back-end blockchain and claim that phishing websites for NFT hacking cannot be regulated using centralized regulations. Centralized regulation can be applied to front-end web pages, such as removing the display of counterfeit NFTs on web pages (of course, the records in the blockchain cannot be deleted) and adding phishing websites to the danger list, etc.} However, as mentioned earlier, the concept of financial scams and unethical NFTs is relatively vague, and most participants do not support centralized regulation. Centralized regulation should only be adopted when the issues harm the interests of the NFT market. Solutions to these issues will be discussed in the next subsection.

\subsubsection{NFT community and Decentralized Autonomy on Ambiguous Issues}
\ 
\newline
\indent
What issues need to apply centralized regulation requires strict analysis, at least with the consent of most community members. Centralized regulations can hardly distinguish financial scams and unethical NFTs. The application of centralized regulation to smart contract vulnerabilities is also controversial~\citep{arroyo2022dao}. Due to the chaos and complexity of NFT market information, the NFT community can play a role in information exchange, dialogue, and project promotion. Similarly, Some studies also point out that NFT creators will form a community for creative communication and rights protection~\citep{sharma2022s}. KoLs and creator communities can help community members identify financial scams and unethical NFTs. These are decentralized autonomous. Decentralized autonomy reflects diversity --- everyone has the right to publish their projects, and the community also has the authority to evaluate and recommend each project. This authority is not tyranny, it will not prevent individuals or teams from publishing works that most community members do not approve of, it will only limit the propagation of such works. 

Decentralized autonomy can also be used in the design of blockchain technology. Blockchain developers are only a small portion of the NFT community, which means that there will inevitably be disagreement between a small number of community managers/blockchain developers and the majority of community members. Therefore, only by fully understanding the needs of community members can researchers develop technologies that are more in line with and maximize the interests of the community. Community members can vote on new technologies or management solutions to determine the solutions that maximize community interests. This voting is a reaction to decentralized autonomy. 

Decentralized autonomy is likely to face a series of issues. The decentralized autonomy of the NFT community is similar to anarcho-communism~\citep{price2008anarchist}. Ideally, each community member acts for community benefits rather than personal benefits. Some scholars call it "crypto-communism"~\citep{husain2020political}. So, it will also face the issues of anarcho-communism. For example, as mentioned in the Twitter analysis, there will be a certain number of KoLs formed in the community, and they may exaggerate certain projects for their benefit rather than community benefits. Community members also need proven incentives to ensure their enthusiasm for participating in decentralized autonomy. Some studies believe that decentralized autonomy often does not rely solely on smart contracts on-chain, but also requires the guarantee of systems and rules off-chain~\citep{de2020blockchain, wang2019decentralized}. Recently, some visual tools can help community members, researchers, and even community regulators better understand decentralized autonomous activities~\citep{arroyo2022dao}.

\rev{NFT community members are also an important part of decentralized autonomy. In an environment where private keys and property cannot be retrieved, it is difficult for centralized regulations to cover all aspects, and NFT community members must protect themselves. Do not trust phishing websites and use standardized smart contracts to reduce the risk of property damage. Of course, members can also choose non-standardized contracts to obtain higher returns, but on the contrary, these contracts may have greater risks or vulnerabilities. Everyone has the freedom to choose and is responsible for their own choices, which is also part of decentralized autonomy.}

\subsubsection{Researching and Promoting New Technologies to Promote Decentralization} 
\ 
\newline
\indent
The development of technology is very important for decentralized autonomy and solving the centralization-decentralization dilemma. With the continuous development of technologies such as AI, some centralization-decentralization dilemmas will be fundamentally resolved. For example, counterfeit NFTs must now pass a centralized manual censor, but with the development of AI technology, deploying AI models in smart contracts may effectively achieve decentralized censor. Solana has lower gas fees compared to ETH and offers more environmentally friendly minting. Ethereum now uses proof-of-stake instead of proof-of-work, which may reduce hashing, and ultimately, gas fees. Some studies have explored hardware security modules (HSM) for private keys to prevent private keys from hacking~\citep{boireau2018securing}.

The above solutions are not mutually exclusive, and some solutions cannot be simply classified as centralized regulation or decentralized autonomy. The DAO event is a good example. In the view of techno-utopians, mandatory rollback is unquestionable centralized regulation. But the actual situation is that whether to accept the rollback is decided by the Ethereum miners. In the end, 90\% of the miners accepted the new agreement to roll back, and Ethereum also had a hard fork. After the hard fork, ETH still maintained rapid development. A small number of community members who opposed centralized regulation were not passively accepted. As P5 said, if community members were dissatisfied with this community, they could establish a new community and determine a new consensus. A small group of community members who were not in favor of forcing transaction rollbacks created ETC through a hard fork and enforced what they called ``purer decentralized autonomy'' on ETC. Of course, they also have to bear the property losses in ETC This is essentially a manifestation of fully guaranteeing everyone's right to choose freely. In this sense, this rollback is an off-chain decentralized autonomy. However, specific solutions need to be effectively formulated according to specific issues. The DAO event was an event that caused panic in the entire community, and the majority of community members united based on common interests. For conventional hacking, it may be difficult to form such a huge force for rollback, and it may eventually require centralized regulation and NFT community members to protect themselves.

\rev{
\subsection{Beyond Digital Collectibles and Bubbles: The Future of NFTs}
Beyond the NFTs market, there is a more fundamental issue, which is the real value of NFTs. On September 20, 2023, a research report showed that the weekly transaction volume of NFTs in July 2023 was 80 million dollars, which was only 3\% of the highest point (August 2021). Of the 73,257 NFT collections the author identified, an eye-watering 69,795 of them have a market cap of 0 Ether (ETH), and many armchair commentators have referred to this crash as ``the death of the NFT''\citep{NFTsale2023}. So what is the uniqueness of NFT, and can this uniqueness help it have a place in the future?

Looking back at the technical features of NFT, we can find that NFT consists of two parts: tokens and assets. Generally speaking, assets are digital files (such as images, videos, games, etc.) and are independent of tokens. In other words, anyone can easily download assets to a personal computer and ``own'' them. The only difference is that this ownership is not recorded on the blockchain. This is an interesting social phenomenon ``right-click mentality''~\citep{bonnet2023impact}. Even though we don’t encourage the free use of digital assets, they can still be sold on centralized platforms. This is why we conduct thought experiments. When the use value is equivalent, the advantage, or the uniqueness of NFT is (technically) decentralized, non-tamperable ownership; while the disadvantage is mainly that the minting process may incur additional costs. Previous research has reported that oligopolistic, centralized digital platforms put creators at a disadvantage~\citep{meier2019rising}. Therefore, technological decentralization at least gives creators new hope~\citep{chalmers2022beyond}. As mentioned in Section 4.3.3, creators receive all the income from selling NFTs without having to share it with the centralized platform. But what cannot be ignored is that the cost of minting NFT is still very high, and it is difficult for unknown individual creators to have enough property to mint NFT on a large scale~\citep{chalmers2022beyond}. These creators run the risk of compromising with big capital, resulting in the emergence of ``centralized capital'' that decentralization supporters do not want to see. What is more critical is the attitude of collectors, which is the market demand. It can be seen from the interview that for most collectors, purchasing NFTs is not for decentralized ownership, but simply for the speculation of the NFTs. If collectors only focus on use value, they will only choose to buy cheaper products --- and this is the disadvantage of NFTs. This shows that although decentralization is a new technology, it is not attractive enough to the people, or in other words, the advantages brought by decentralization are not enough to offset its disadvantages. This also seems to explain the rapid shrinkage of the NFT market.

However, this does not mean that NFTs are worthless. Viewed through a historical lens, bubbles can sometimes create social benefits by channeling capital into more dynamic economic sectors, such as railway companies in the 1840s, bicycle companies in the 1890s, or internet firms in the late 1990s.~\citep{chalmers2022beyond} Many scholars and experts still think about the practical application scenarios of NFTs like real estate, digital identity, and Token-gated access~\citep{NFTsale2023}. Unfortunately, it remains to be seen whether decentralization performs better than centralization in these scenarios. A simple example is that if your car recorded on the blockchain is stolen, you should still seek help from the police station (which is a centralized agency) rather than the decentralized organization. So the future of NFTs remains unclear, it seems not easy to find real application scenarios.
}

\subsection{Limitations}
Our study has several limitations. First of all, our interview sample size is relatively small (N = 15) and only two types of stakeholders were included in the sample, while the latest research points out that there may be six categories of stakeholders~\citep{baytacs2022stakeholders}. Interviews with other stakeholders can be conducted in the future to obtain more comprehensive information and perspectives. Secondly, our lack of understanding of the literature on the social and economic aspects of blockchain may make us lack a specific perspective when expressing our views. So we can only make policy and design recommendations from a technical perspective. Thirdly, some research team members are not members of the NFT community. Therefore, there may be issues that we ignore in the context of the centralization-decentralization dilemma. 
Finally, due to the limited number of tweets, we cannot capture many people's views, especially when a lot of people tend to post their opinions on other social platforms or blogs.

\section{Conclusion}
We use a mix of social media analysis and interviews to study the perceptions and concerns of stakeholders in the NFT market. Compared with prior research on the NFT market and the entire blockchain industry, our research has the following three contributions: first, through the combination of qualitative and quantitative analysis, we identify the issues that users are most concerned about in the context of the centralization-decentralization dilemma; second, we learn about perceptions of different stakeholders (i.e., creators, collectors) in the NFT market on the centralization-decentralization dilemma; third, we propose design and policy implications to solve the centralization-decentralization dilemma in the NFT market. \rev{Finally, We think about the real value of NFT.} We believe that research in the NFT market and the entire blockchain field will be an emerging and important direction.
\begin{acks}
We express our gratitude to the anonymous reviewers at CSCW (Conference on Computer Supported Cooperative Work) 2024 for their insightful feedback, which has substantially contributed to the refinement and advancement of our research for publication. X. Tong and L. Zhang acknowledge the Data Science Research Center at Duke Kunshan University for funding their project entitled “A Data-Centered Approach to Investigate the Ethics Issues on Non-Fungible Tokens (NFT): Trust, Privacy, Transparency, and Fairness” under the Data + X Program. Luyao Zhang is supported by National Science Foundation China on the project entitled “Trust Mechanism Design on Blockchain: An Interdisciplinary Approach of Game Theory, Reinforcement Learning, and Human-AI Interactions (Grant No. 12201266).” 
\end{acks}
\bibliographystyle{ACM-Reference-Format}
\bibliography{acmart}


\section{Appendix}
\beginsupplement
\subsection{Interview Questions}

(1) Can you briefly introduce yourself (add details, e.g., age, job, region/country, years with NFT and cybermoney, etc.) and tell us about your prior experiences with NFT?

\textbf{NFT artwork creation}(for creators)

(2) What was your motivation for creating NFT artworks/collections?

(3) Would you adjust your idea about creation according to the market needs? Why or why not?

(4) What is one of your most satisfying NFT works and what is your most popular (highest priced) NFT work? Are they the same?

\textbf{Participate in art trading}

(5) Have you ever participated in any offline art trading(e.g. Art auction)? If so, what is it about?

(6) What are the similarities and differences between offline art trading and online trading, more specifically, between offline art trading and NFT trading?

(7)Why did you choose to trade your artwork at the NFT platforms instead of the conventional digital art trading platforms? (e.g. ArtFire, Artplode, Saatchi Art, skeb, Instagram, etc.)

(8) What do you think are the similarities and differences between NFT and conventional digital art trading platforms?
 
(9) What latest trends do you follow in the NFT market (e.g. an overpriced NFT, or NFT news, etc.)? How do you find out the information(e.g. transaction information) (directly on Opensea/on Twitter or other)?

\textbf{Hacking and privacy}

(10) What do you know about the NFT’s smart contract and its terms?
 
(11) When there is a problem with smart contracts, threatening the property or privacy of many people, do you think it should take centralised regulation(or is codes law)?

(12) Who do you think should be responsible for the transaction security for consumers who purchase your NFT artwork? Why do you think so?

(13) What do you know about the trading principles of NFT(e.g. Why can NFT ownership not be changed, the association of cybermoney with fiat currency may expose personal information)? Can you explain?

(14) Do you have any concerns about your personal information(e.g. Personal identity and wallet binding) being exposed during this transaction? Why or why not?

(15) Have you ever heard of (or experienced) “theft” of an NFT trading account?
If no, explain the theft briefly and then ask.
If yes, (Considering that it is not retrievable) does the theft affect your motivation to purchase NFT? Why or why not?
What improvements(e.g. Participate in community governance) would you like to see to help improve the security of the NFT transactions?

\textbf{Fairness, Bias, transparency, and  trust}

(16) Have you ever heard of (or purchased) the NFT project with minor modifications being sold as a new NFT project?

(17) What do you think about this way of producing NFT? Should this be regulated(e.g. by the government)? Why or why not?

(18) If your NFT artwork is being copied, how would you defend your interests?

(19) (For creators)What ethical problems or concerns (diversity, fairness, trust, bias etc.) do you have about NFT artwork/collections, either yours or other creators’? 
(If Yes)What implications or inspirations do you get from these ethical problems (diversity, ethics, fairness, trust, bias etc.) when creating your NFT artwork?

(20) Have you heard about any NFT controversial or negative news? (e.g. Bored Ape suspected of Nazi connection, Meta slave involved in racism).
What other similar controversial NFTs do you know of?

(21) Compared to conventional digital art trading platforms, What are the new features of ethical issues in the NFT market?

(22) Everyone has a different code of ethics. Do you think it's reasonable to report a work you like and take it down because most people don't like it? 
If yes, How are minority rights protected? Does this affect diversity? Does this go against the idea of decentralization?
If no, How to solve this ethical dilemma?

(23) In the case of the above NFT , who do you think is responsible? Under what circumstances are others(besides the creators) responsible for NFT ethics issues?

(24) Do you know the NFT’s transaction gas on Ethereum? 
Does this affect your enthusiasm for creating NFT? 
Do you think this will affect the promotion of the NFT?

(25) Some people criticize decentralization, or mining, for wasting electricity and environmental problems, what do you think about that?

(26) Many people now use virtual currencies and NFTs as wealth management products. Does this run counter to the idea of decentralization?

(27) Do you think there are inequalities in the current nft art market? If so,Which aspect of inequality(e.g.diversity, ethics,fairness, trust, bias etc.) is it? What are the reasons for inequality?

(28) Considering that many NFTs can be used for non-commercial use for free (separation of access and ownership), will this affect consumers' enthusiasm for buying?

(29) Can NFTs be used for physical objects?

\textbf{Finance, Security,  Fraud and Ponzi Schemes}

(30) According to your observation, which types of NFTs are more popular?What are your thoughts on the more popular NFT projects currently on the market (e.g. Bored Ape, etc.)?
What are the reasons for its popularity?

(31) Do you think these NFTs have a price commensurate with their ( artistic ) value?

(32) Do you promote your NFT artwork on social media? If so, how do you promote them? Do you promote your NFT artwork to friends and around you? If so, how do you promote it? What is your most successful promotion strategy? Why does it work (or not)?

(33) Do you think that NFT artwork has potential financial risks? Why or why not?

\section{Quantitative Analysis}
\subsection{Opinion leaders}
\begin{table}[H]
\caption{Top 20 nodes (part of opinion leaders) according to Eigencentrality, with attributes of followersCount, eccentricity, betweenesscentrality, degree, indegree and outdegree.}
\label{tab:top20nodes}
\begin{tabular}{@{}llllllll@{}}
\toprule
Id & followersCount & Eccentricity & betweenesscentrality & eigencentrality & degree & indegree & outdegree \\ \midrule
NFTethics        & 65161891 & 18 & 0.060081 & 1        & 7294 & 7162 & 132 \\
diverse          & 3087285  & 19 & 0.005431 & 0.180818 & 1244 & 1240 & 4   \\
opensea          & 9910610  & 0  & 0        & 0.168713 & 1226 & 1226 & 0   \\
psychedelic\_nft & 949795   & 0  & 0        & 0.166322 & 1211 & 1211 & 0   \\
BAYC2745         & 3247690  & 21 & 0.003368 & 0.127838 & 949  & 940  & 9   \\
Coinbase\_NFT    & 2161621  & 22 & 0.002424 & 0.068506 & 499  & 497  & 2   \\
zachxbt          & 8116342  & 19 & 0.004041 & 0.062464 & 347  & 327  & 20  \\
beaniemaxi       & 4351131  & 19 & 0.001792 & 0.056376 & 310  & 295  & 15  \\
rugpullfinder    & 1495774  & 18 & 0.002156 & 0.050508 & 277  & 258  & 19  \\
worldofwomennft  & 827191   & 21 & 0.004522 & 0.048845 & 334  & 333  & 1   \\
farokh           & 5996892  & 19 & 0.000865 & 0.043527 & 209  & 200  & 9   \\
binance          & 40166534 & 1  & 0.000041 & 0.04342  & 219  & 217  & 2   \\
bapesclan        & 86216    & 0  & 0        & 0.043317 & 203  & 203  & 0   \\
osf\_nft         & 11223003 & 19 & 0.00195  & 0.040161 & 208  & 184  & 24  \\
NFTherder        & 1908205  & 18 & 0.002609 & 0.038722 & 178  & 156  & 22  \\
betty\_nft       & 6261677  & 18 & 0.003327 & 0.0372   & 302  & 268  & 34  \\
Zeneca\_33       & 3755295  & 19 & 0.00193  & 0.034547 & 157  & 151  & 6   \\
cerealclubnft    & 220774   & 0  & 0        & 0.03275  & 135  & 135  & 0   \\
ABigThingBadly   & 2080338  & 19 & 0.00468  & 0.031127 & 149  & 119  & 30  \\
TheSandboxGame   & 3214768  & 22 & 0.000581 & 0.030981 & 132  & 127  & 5   \\ \bottomrule
\end{tabular}
\end{table}

\subsection{Biagrams}
\begin{table}[H]
\centering
\caption{Top 20 bigrams in terms of five main concerned aspects.}
\label{tab:biagram-table}
\resizebox{\textwidth}{!}{\begin{tabular}{llllllllll}
\textbf{Trust}             & \textbf{Count} & \textbf{Bias}             & \textbf{Count} & \textbf{Diverse}       & \textbf{Count} & \textbf{Fairness}   & \textbf{Count} & \textbf{Ethics}       & \textbf{Count}  \\
(trust, wallet)            & 7619           & (nft, space)              & 176            & (nft, space)           & 1074           & (moral, story)      & 319            & (monthly, payment)    & 2108            \\
(strong, team)             & 6536           & (nft, project)            & 145            & (nft, project)         & 788            & (nft, space)        & 240            & (home, equity)        & 1506            \\
(transparent, road)        & 3670           & (confirmation, bias)      & 127            & (nft, community)       & 475            & (nft, project)      & 138            & (debt, learn)         & 1040            \\
(predictable, transparent) & 3629           & (nft, collection)         & 87             & (diverse, inclusive)   & 462            & (moral, compass)    & 129            & (free, ebook)         & 1016            \\
(road, map)                & 3575           & (nft, community)          & 69             & (inclusive, community) & 438            & (ethical, nft)      & 85             & (get, cash)           & 987             \\
(project, strong)          & 3523           & (bias, towards)           & 65             & (nft, collection)      & 407            & (moral, support)    & 73             & (equity, agreement)   & 735             \\
(team, predictable)        & 3430           & (survivorship, bias)      & 62             & (diverse, nft)         & 398            & (nft, community)    & 69             & (additional, debt)    & 727             \\
(team, transparent)        & 3374           & (might, bias)             & 43             & (nft, artist)          & 306            & (ethical, issue)    & 58             & (cash, home)          & 722             \\
(map, planned)             & 3110           & (little, bias)            & 42             & (inclusive, nft)       & 282            & (moral, high)       & 57             & (reverse, mortgage)   & 675             \\
(planned, projected)       & 3058           & (bias, nft)               & 40             & (diverse, community)   & 282            & (high, ground)      & 50             & (heloc, reverse)      & 674             \\
(project, future)          & 2985           & (unit, bias)              & 39             & (would, love)          & 269            & (nft, ethics)       & 47             & (loan, heloc)         & 674             \\
(good, project)            & 2931           & (without, bias)           & 37             & (diverse, metaverse)   & 214            & (make, money)       & 46             & (agreement, quantmre) & 674             \\
(excellent, project)       & 2913           & (personal, bias)          & 36             & (would, like)          & 199            & (feel, like)        & 46             & (payment, added)      & 674             \\
(project, roadmap)         & 2891           & (bit, bias)               & 36             & (web, community)       & 176            & (ethical, way)      & 45             & (added, debt)         & 674             \\
(future, strong)           & 2891           & (racism, homophobia)      & 35             & (inclusive, diverse)   & 167            & (nft, ethic)        & 41             & (mortgage, monthly)   & 674             \\
(transparent, planned)     & 2783           & (culture, discrimination) & 33             & (community, would)     & 156            & (moral, ethical)    & 38             & (quantmre, loan)      & 674             \\
(planned, project)         & 2721           & (bias, think)             & 33             & (trying, build)        & 154            & (nft, art)          & 37             & (learn, free)         & 533             \\
(trust, project)           & 2652           & (history, culture)        & 32             & (diverse, group)       & 154            & (twitter, account)  & 37             & (payment, learn)      & 501             \\
(nft, project)             & 2483           & (embracing, diversity)    & 32             & (nft, art)             & 154            & (many, people)      & 36             & (tap, home)           & 483             \\
(near, future)             & 2413           & (symbol, arts)            & 32             & (love, join)           & 151            & (moral, obligation) & 36             & (download, free)      & 441            
\end{tabular}}
\end{table}

\subsection{Topic Modeling}
\label{app:topic_modeling}
\subsubsection{Topics of Tweets from KOLs}
\label{app:topics_KOL}
Top 10 words for topic \#0:\newline
['project', 'transparent', 'space', 'community', 'people', 'diverse', 'inclusive', 'nft', 'trust', 'eth']

Top 10 words for topic \#1:\newline
['diverse', 'transparent', 'metaverse', 'project', 'eth', 'space', 'community', 'people', 'nft', 'trust']

Top 10 words for topic \#2:\newline
['diverse', 'metaverse', 'trust', 'inclusive', 'community', 'nft', 'space', 'people', 'transparent', 'project']

Top 10 words for topic \#3:\newline
['inclusive', 'metaverse', 'eth', 'community', 'diverse', 'space', 'people', 'trust', 'nft', 'project']

Top 10 words for topic \#4:\newline
['transparent', 'space', 'trust', 'community', 'people', 'diverse', 'inclusive', 'nft', 'project', 'eth']

Top 10 words for topic \#5:\newline
['metaverse', 'eth', 'trust', 'project', 'diverse', 'community', 'people', 'inclusive', 'nft', 'space']

Top 10 words for topic \#6:\newline
['project', 'eth', 'space', 'people', 'trust', 'inclusive', 'community', 'diverse', 'transparent', 'nft']

Top 10 words for topic \#7:\newline
['eth', 'inclusive', 'diverse', 'community', 'transparent', 'trust', 'space', 'project', 'nft', 'people']

Top 10 words for topic \#8:\newline
['transparent', 'trust', 'people', 'project', 'eth', 'community', 'space', 'nft', 'inclusive', 'diverse']

Top 10 words for topic \#9:\newline
['community', 'metaverse', 'transparent', 'inclusive', 'trust', 'project', 'people', 'diverse', 'nft', 'space']

Top 10 words for topic \#10:\newline
['eth', 'project', 'people', 'space', 'diverse', 'trust', 'transparent', 'inclusive', 'nft', 'community']

Top 10 words for topic \#11:\newline
['trust', 'diverse', 'eth', 'people', 'project', 'metaverse', 'nft', 'community', 'space', 'inclusive']

Top 10 words for topic \#12:\newline
['metaverse', 'eth', 'trust', 'community', 'diverse', 'project', 'inclusive', 'people', 'nft', 'space']

Top 10 words for topic \#13:\newline
['people', 'transparent', 'trust', 'project', 'community', 'space', 'eth', 'diverse', 'nft', 'metaverse']

Top 10 words for topic \#14:\newline
['inclusive', 'diverse', 'transparent', 'eth', 'space', 'people', 'nft', 'trust', 'community', 'project']

\subsubsection{Topics of Tweets from Ordinary Users}
\label{app:topics_Od}
Top 10 words for topic \#0:\newline
['space', 'super', 'building', 'utility', 'transparency', 'team', 'transparent', 'nft', 'project', 'community']

Top 10 words for topic \#1:\newline
['possible', 'trust', 'nft', 'doesnt', 'seen', 'read', 'sell', 'price', 'ive', 'bias']

Top 10 words for topic \#2:\newline
['good', 'future', 'success', 'impresive', 'successfully', 'thanks', 'believe', 'great', 'trust', 'project']

Top 10 words for topic \#3:\newline
['bought', 'friend', 'year', 'month', 'happy', 'thank', 'life', 'day', 'nft', 'trust']

Top 10 words for topic \#4:\newline
['little', 'tweet', 'problem', 'time', 'smart', 'contract', 'right', 'nft', 'trust', 'got']

Top 10 words for topic \#5:\newline
['transparent', 'investment', 'market', 'platform', 'company', 'crypto', 'asset', 'token', 'equity', 'nft']

Top 10 words for topic \#6:\newline
['planned', 'projected', 'map', 'predictable', 'strong', 'team', 'road', 'transparent', 'good', 'project']

Top 10 words for topic \#7:\newline
['got', 'lost', 'account', 'metamask', 'nft', 'hacked', 'help', 'trust', 'need', 'wallet']

Top 10 words for topic \#8:\newline
['waiting', 'successful', 'planned', 'excellent', 'strong', 'roadmap', 'team', 'transparent', 'future', 'project']

Top 10 words for topic \#9:\newline
['miss', 'eth', 'check', 'sol', 'new', 'want', 'follow', 'dont', 'nft', 'trust']

Top 10 words for topic \#10:\newline
['look', 'yes', 'try', 'bro', 'good', 'nft', 'know', 'people', 'process', 'trust']

Top 10 words for topic \#11:\newline
['make', 'buying', 'project', 'nft', 'like', 'money', 'market', 'better', 'lot', 'trust']

Top 10 words for topic \#12:\newline
['hey', 'people', 'diverse', 'build', 'web', 'space', 'inclusive', 'nft', 'community', 'love']

Top 10 words for topic \#13:\newline
['info', 'value', 'appreciate', 'let', 'lets', 'ethical', 'story', 'moral', 'nft', 'transparency']

Top 10 words for topic \#14:\newline
['coinbase', 'wallet', 'nft', 'trust', 'meta', 'btc', 'crypto', 'opensea', 'logo', 'eth']

Top 10 words for topic \#15:\newline
['mind', 'piece', 'hold', 'people', 'world', 'trust', 'work', 'nft', 'artist', 'art']

Top 10 words for topic \#16:\newline
['matter', 'today', 'player', 'wait', 'nft', 'game', 'sale', 'fairness', 'soon', 'coming']

Top 10 words for topic \#17:\newline
['thing', 'twitter', 'hope', 'lol', 'space', 'ethic', 'trust', 'like', 'people', 'nft']

Top 10 words for topic \#18:\newline
['opportunity', 'solid', 'plan', 'transparent', 'great', 'amazing', 'trust', 'big', 'project', 'team']

Top 10 words for topic \#19:\newline
['space', 'gender', 'inclusivity', 'play', 'thank', 'diversity', 'equality', 'woman', 'nft', 'game']

Top 10 words for topic \#20:\newline
['space', 'real', 'nft', 'question', 'theyre', 'thread', 'like', 'trust', 'people', 'guy']

Top 10 words for topic \#21:\newline
['inclusive', 'drop', 'metaverse', 'new', 'set', 'unique', 'world', 'nft', 'diverse', 'collection']

Top 10 words for topic \#22:\newline
['point', 'thing', 'nft', 'mean', 'said', 'scam', 'rug', 'say', 'didnt', 'project']

Top 10 words for topic \#23:\newline
['term', 'giveaway', 'want', 'nft', 'trust', 'free', 'long', 'join', 'discord', 'mint']

Top 10 words for topic \#24:\newline
['technology', 'creator', 'using', 'transparent', 'platform', 'buy', 'use', 'blockchain', 'trust', 'nft']

\subsection{Thematic Analysis}
\href{https://docs.google.com/spreadsheets/d/1mePp4MDCZb3UsiV_byC9oceRaZ3fZQxG/edit?usp=sharing&ouid=101176736800945590817&rtpof=true&sd=true}{Link to the tweet thematic doing book.}
\label{codebook}

\end{document}